\documentclass[twocolumn,english,aps,prl,superscriptaddress,floats,nobibnotes,longbibliography]{revtex4-1}

\usepackage[utf8]{inputenc}
\usepackage{amsmath}
\usepackage{amssymb,bm}
\usepackage{amsthm}
\usepackage{setspace}
\usepackage{graphicx}
\usepackage{braket}
\usepackage{mathrsfs}
\usepackage{float}
\usepackage[colorlinks = true,linkcolor = blue,urlcolor  = blue,citecolor = blue,anchorcolor = blue]{hyperref}
\usepackage[english]{babel}
\usepackage{bm}
\usepackage{csquotes}
\usepackage{hyperref}

\begin{document}

\title{Theory of anisotropic magnetotransport beyond the constant relaxation-time approximation in multi-Weyl semimetals}
\author{Azaz Ahmad}
\affiliation{Department of Physics, Indian Institute of Technology Bombay, Powai, Mumbai 400076, India.}
\author{Shantanu Mukherjee}
\affiliation{Department of Physics, Indian Institute of Technology Bombay, Powai, Mumbai 400076, India.}
\author{Mohd.~Hashim Raza}
\affiliation{Maharani Lal Kunwari College, Tulsipur Road, Balrampur, Uttar Pradesh-271201, India.}
\author{Gautham Varma K.}
\affiliation{Department of Physics, Indian Institute of Technology Delhi, Hauz Khas, New Delhi 110016, India.}
\author{Takami Tohyama}
\affiliation{Department of Applied Physics, Tokyo University of Science, Katsushika, Tokyo 125-8585, Japan.}
\vspace{1cm}
\date{\today}
\begin{abstract}
Multi-Weyl semimetals, characterized by Weyl nodes carrying higher monopole charges, provide a platform for exploring topological magnetotransport beyond the conventional Weyl paradigm. Here, we develop a generalized semiclassical Boltzmann framework for multi-Weyl semimetals with arbitrary monopole charge $N$, incorporating the orbital magnetic moment (OMM) into the quasiparticle dispersion and Berry-curvature-modified dynamics, together with a momentum-dependent collision integral accounting for intra- and intervalley scattering. We investigate the longitudinal magnetoconductivity (LMC) and planar Hall conductivity (PHC) for $N=1$, $2$, and $3$, considering both nonmagnetic and magnetic impurity scattering. In the absence of OMM, the LMC remains positive and quadratic in the magnetic field throughout the investigated scattering regime, consistent with conventional chiral-anomaly-assisted transport. Including OMM, however, leads to a sign reversal of the LMC under sufficiently strong intervalley scattering for the $\sigma_0$ and $\sigma_z$ impurity channels. Remarkably, the critical intervalley scattering strength decreases systematically with increasing monopole charge, indicating an enhanced tendency toward negative LMC in higher-charge Weyl systems. Furthermore, we find that the PHC also undergoes a sign reversal for $N=2$ and $N=3$ in the same impurity channels, while remaining positive for $N=1$ within the investigated parameter range. The angular dependences retain their characteristic $\sin^2\gamma$-like and $\sin2\gamma$ profiles for LMC and PHC, respectively, even across the sign reversal, with a small additional contribution to the LMC at perpendicular field orientations. Our results demonstrate that the interplay of Berry curvature, OMM, monopole charge, and impurity scattering can produce qualitatively distinct weak-field magnetotransport responses beyond the conventional chiral-anomaly picture.
\end{abstract}
\maketitle
\section{Introduction}
\label{sec:Introduction}
Three-dimensional topological quantum materials have attracted considerable attention owing to their unconventional electronic properties and potential applications in electronic and spintronic devices. Among them, Weyl semimetals constitute an important class of topological materials, hosting low-energy quasiparticles that behave as massless Weyl fermions near isolated band crossings in momentum space~\cite{bevan1997momentum,wan2011topological,Yan_2017,armitage2018weyl,burkov2011weyl}. Each Weyl node acts as a source or sink of Berry curvature and carries a quantized monopole charge, whose sign determines its chirality. This nontrivial band topology gives rise to remarkable phenomena, including Fermi-arc surface states, the anomalous Hall effect, and chiral-anomaly-related transport responses~\cite{adler1969axial,nielsen1981no,son2012berry,goswami2015optical,nandy2017chiralplanar}. While conventional Weyl nodes carry unit monopole charge ($N=1$) and exhibit linear dispersion in all momentum directions, multi-Weyl semimetals support higher-charge nodes ($N=2,3$) protected by crystalline rotational symmetries. These higher-order nodes exhibit quadratic or cubic dispersion in the transverse plane while retaining linear dispersion along the symmetry axis, resulting in anisotropic Berry curvature, modified density of states, and unconventional magnetotransport properties~\cite{fang2012multiweyl}.

The existence and arrangement of Weyl nodes are constrained by crystal symmetries and lattice topology. In three-dimensional crystals, Weyl nodes require the breaking of either time-reversal symmetry (TRS) or inversion symmetry (IS). Furthermore, the Nielsen--Ninomiya no-go theorem requires the total monopole charge within the Brillouin zone to vanish, ensuring the presence of nodes with compensating chiralities~\cite{nielsen1981no,nielsen1983adler}. The topological charge of each Weyl node is characterized by a quantized Chern number, defined through the Berry-curvature flux over a closed surface enclosing the node. This nontrivial band topology underlies various anomalous electromagnetic and transport responses, whose magnitudes also depend on the band structure and scattering mechanisms~\cite{xiong2015evidence,nagaosa2010anomalous,sharma2016nernst,zyuzin2017magnetotransport,kim2014boltzmann}.

Beyond conventional single-Weyl systems, higher-charge Weyl nodes have been extensively investigated theoretically and realized in engineered photonic structures~\cite{chen2016photonic}. The anisotropic dispersion of multi-Weyl fermions, which is linear along the symmetry axis and quadratic or cubic in the transverse plane for $N=2$ and $N=3$, respectively, gives rise to distinctive electronic and transport properties. In particular, the monopole charge and anisotropic band structure govern characteristic scaling behaviors in the density of states, optical conductivity, and semiclassical transport coefficients~\cite{ahn2017optical,park2017semiclassical,zivieri2023high}. Their consequences have also been explored in thermoelectric, nonlinear, and anomalous transport responses~\cite{nag2020thermoelectric,nag2022distinct,zeng2021nonlinear,xiong2022understanding,gorbar2017anomalous,dantas2020non}. These findings establish multi-Weyl semimetals as a platform for investigating how higher topological charge and anisotropic band dispersion influence electronic transport.

The higher monopole charge and anisotropic dispersion of multi-Weyl nodes modify the Berry-curvature distribution and consequently influence semiclassical carrier dynamics and magnetotransport~\cite{park2017semiclassical,dantas2018magnetotransport}. An important mechanism underlying these responses is the chiral anomaly, which describes the nonconservation of the particle number associated with individual Weyl nodes in the presence of electric and magnetic fields satisfying $\mathbf{E}\cdot\mathbf{B}\neq0$, while the total electric charge remains conserved~\cite{son2013chiral}. In the weak-field semiclassical regime, the anomaly can generate positive longitudinal magnetoconductivity, corresponding to negative longitudinal magnetoresistance, and an associated planar Hall response~\cite{son2013chiral,nandy2017chiralplanar}. The interplay of Berry curvature, monopole charge, anisotropic band dispersion, and scattering processes therefore plays a central role in determining magnetotransport in multi-Weyl semimetals.

Recent theoretical studies have further demonstrated that the electrical, thermal, and thermoelectric responses of multi-Weyl semimetals exhibit nontrivial dependencies on monopole charge, temperature, and anisotropic band dispersion~\cite{nag2020thermoelectric,ghosh2024electricthermoelectric,medel2024electric}. In coplanar field configurations, the longitudinal magnetoconductivity (LMC) and planar Hall conductivity (PHC) exhibit characteristic magnetic-field and angular dependences, while the associated thermal and thermoelectric responses are also sensitive to the topological charge and field orientation~\cite{dantas2018magnetotransport,ghosh2024electricthermoelectric}. Furthermore, the interplay of Berry curvature and orbital magnetic moment (OMM) can significantly modify the longitudinal and transverse transport coefficients, emphasizing the importance of band-geometric corrections in multi-Weyl systems~\cite{medel2024electric}.

The stability of higher-charge Weyl nodes is closely tied to crystalline rotational symmetry. Double-Weyl nodes can be protected by $C_4$ or $C_6$ symmetry, whereas triple-Weyl nodes require $C_6$ symmetry within the conventional two-band classification~\cite{fang2012multiweyl}. Breaking the protecting rotational symmetry can split a multi-Weyl node into several single-Weyl nodes, redistributing its monopole charge while preserving the total topological charge~\cite{fang2012multiweyl}. Such symmetry constraints are important for identifying the regime in which higher-charge quasiparticles and their associated transport properties can be realized, complementing the established studies of Berry-curvature-mediated magnetotransport in Weyl systems~\cite{burkov2014anomalous,son2013chiral,kim2014boltzmann,sharma2016nernst}.

In the present work, we investigate the longitudinal magnetoconductivity (LMC) and planar Hall conductivity (PHC) of multi-Weyl semimetals with monopole charges $N=1$, $2$, and $3$, incorporating Berry curvature, orbital magnetic moment (OMM), and momentum-dependent intra- and intervalley impurity scattering within the semiclassical Boltzmann framework beyond the constant relaxation-time approximation. We systematically examine how these effects, together with the nature of magnetic and nonmagnetic impurities, influence the magnitude, sign, and angular dependence of the magnetotransport response. The remainder of the manuscript is organized as follows. In Sec.~\ref{sec:double_weyl}, we introduce the multi-Weyl Hamiltonian and derive the corresponding Berry curvature and OMM. In Sec.~\ref{sec:transport_DW}, we formulate the semiclassical Boltzmann transport theory with a momentum-dependent collision integral and particle-number conservation. In Sec.~\ref{Sec.:Results and discussion}, we present the numerical results for LMC and PHC, focusing on their magnetic-field and angular dependences, the influence of intervalley scattering, and the emergence of sign reversals for different monopole charges and impurity channels. Finally, Sec.~\ref{Sec:Conclusions} summarizes our main findings, while the Appendices provide the technical derivations and analytical checks supporting the numerical calculations.

\section{Multi Weyl semimetals}
\label{sec:double_weyl}

The low-energy Hamiltonian describing a multi Weyl node (topological charge $C=\pm N$) can be written within a minimal two-band model as~\cite{jian2015correlated,chen2016thermoelectric,huang2016new,singh2018tunable,bradlyn2016beyond,bai2026single,li2025double,medel2024electric,medel2026nonperturbative,medel2024publisher,nag2021magneto}:
\begin{equation}
H^{\chi}_N(\mathbf{k}) = \chi \alpha_{N} k^N_{\bot} \left[ \cos \left(N \phi_{k} \right) 
\sigma_{x} +\sin \left( N \phi_{k} \right) \sigma_{y} \right] + \chi \hbar v k_z \sigma_{z},
\label{eq:H_MW}
\end{equation}
where $\chi = \pm 1$ denotes the chirality of the node,~$k_\perp^2 = k_x^2 + k_y^2$, $v$ is the Fermi velocity along the $z$-direction, $\boldsymbol{\sigma} = (\sigma_x,\sigma_y,\sigma_z)$ are the Pauli matrices, and $\mathbf{k} = k ~(\sin\theta\cos\phi,\sin\theta\sin\phi,\cos\theta)$ is measured from the nodal point. It is important to emphasize that the parameter $\alpha_N$ carries nontrivial dimensional dependence on $N$. In particular, $\alpha_1$ has the dimension of velocity, whereas $\alpha_2$ has the dimension of inverse mass. More generally, for $N>2$, $\alpha_N$ necessarily incorporates an additional microscopic scale (e.g., a momentum cutoff or lattice scale) in order to ensure that the Hamiltonian retains the correct dimension of energy. This reflects the fact that {\color{black}higher-order Weyl nodes are not scale invariant} and require an intrinsic ultraviolet scale. More specifically, $\alpha_N \in \{\hbar v_F, \hbar^2/2m, m v_F^2/\lambda^3 \}$, with $\lambda$ has dimensions of inverse length and introduced to parametrize the cubic dispersion coefficient; it is not identified with the ultraviolet cutoff of the continuum model.

The Hamiltonian in Eq.~(\ref{eq:H_MW}) can be written in a compact form as
 with
\begin{align}
& H=\mathbf{d}\cdot \boldsymbol{\sigma} \label{Hamiltonian compact},\\ &\mathbf{d}=\left( \alpha_{N} k^N_{\bot}\cos \left( N \phi_{k} \right), 
\alpha_{N} k^N_{\bot}\sin \left(N \phi_{k} \right), \hbar v k_z \right)\nonumber.
\end{align}

The corresponding energy dispersion is given by
\begin{equation}
\epsilon_{\mathbf{k}}^{\pm} = \pm \sqrt{ \alpha^2_{N} k^{2 N}_{\bot} + \hbar ^2v^2 k^2_z}= \pm |\mathbf{d}|,
\label{eq_multi2_new}
\end{equation}
where $\pm$ denote the conduction and valence bands, respectively. Since our focus is to study the transport properties in the zero-temperature limit and we have taken the Fermi energy in the conduction band, we will drop the irrelevant band index. It is evident from
Eq.~(\ref{eq_multi2_new}) that the monopole charge $N$ controls both the topological
structure of the Bloch states and the anisotropy of the dispersion.

For $N=1$ with $\alpha_1=\hbar v$, one recovers the conventional isotropic Weyl dispersion
$\epsilon_{\mathbf{k}}=v\hbar \sqrt{k_x^2+k_y^2+k_z^2}$. In contrast, for $N>1$, the dispersion
becomes anisotropic: for $N=2$ it is quadratic in $k_x$ and $k_y$ while remaining linear in $k_z$,
and for $N=3$ it becomes cubic in the transverse plane and linear along the $z$-direction.
Throughout this work, we restrict ourselves to type-I multi-Weyl semimetals, where the nodes are untilted and well separated in momentum space.

To determine the eigen vectors and Berry curvature of multi-Weyl system, we now parametrize $\hat{d}$ in terms of it's angles $\theta_d,\,\phi_d$ as
\begin{equation}
    \mathbf{d} = |\mathbf{d}| (\sin\theta_d \cos\phi_d, \sin\theta_d\sin\phi_d, \cos\theta_d).
\end{equation}
The compact form of the Hamiltonian in Eq~\eqref{Hamiltonian compact} along with the above parametrized form of $\mathbf{d}$ allows us to write the Bloch eigen states corresponding to $\chi=\pm 1$ in terms of the parameters $\theta_d,\,\phi_d$ introduced above as
\begin{align}
|u^{+}\rangle =
\begin{pmatrix}
\cos(\theta_d/2) e^{-i\phi_d} \\
\sin(\theta_d/2)
\end{pmatrix}, \quad
|u^{-}\rangle =
\begin{pmatrix}
\sin(\theta_d/2) e^{-i\phi_d} \\
- \cos(\theta_d/2)
\end{pmatrix},
\label{eq:eigenstates_DW}
\end{align}
where we have used the following relations
\begin{equation}
    \cos\theta_d = \frac{\hbar v_f k_z}{|\mathbf{d}|},\, \alpha_N k_\perp^N =|\mathbf{d}| \sin\theta_d,\, \phi_d = N\phi.
\end{equation}
The winding number N in the azimuthal phase ($N\phi$) directly reflects the topological charge $C=\pm N$. \textcolor{black}{For a two band model, the Berry curvature can also be written as $\Omega_{\mu\nu}= \hat{d}\cdot (\partial_{k_\mu}\hat{d} \times \partial_{k_\nu}\hat{d} )$. This will allow us to write the expression of Berry Curvature in terms of these angles $\theta_d,\,\phi_d$. To proceed we write $\partial_{k_\mu}$ using chain rule
\begin{align}
    \partial_{k_\mu} = \partial_{\theta_d}\,\, \partial_{k_\mu}\theta_d + \partial_{\phi_d} \,\,\partial_{k_\mu}\phi_d
\end{align}
This will lead us to the following expression of the Berry curvature
\begin{equation}\label{Berry3}
    \Omega_{\mu\nu}= \frac{\hat{d}\cdot (\partial_{\theta_d}\hat{d} \times \partial_{\phi_d}\hat{d} )}{2} \left(\partial_{k_\mu}\theta_d \,\partial_{k_\nu}\phi_d - \partial_{k_\mu}\phi_d \,\partial_{k_\nu}\theta_d\right).
\end{equation}
Where the expression $\hat{d}\cdot (\partial_{\theta_d}\hat{d} \times \partial_{\phi_d}\hat{d} )$ is defined purely in space of $\hat{d}$ and is given by
\begin{equation}
    \hat{d}\cdot (\partial_{\theta_d}\hat{d} \times \partial_{\phi_d}\hat{d} ) = \sin\theta_d
\end{equation}
With this we get the useful expression of $\mathbf{\Omega}$ as
\begin{equation}\label{Berry vecor}
    \Omega_\lambda= \frac{1}{2} \sin\theta_d\, \varepsilon_{\lambda \mu \nu}\left(\partial_{k_\mu}\theta_d \,\partial_{k_\nu}\phi_d - \partial_{k_\mu}\phi_d \,\partial_{k_\nu}\theta_d\right).
\end{equation}
Using these expressions, we evaluate the Berry curvature vector for the conduction band, given below
\begin{equation}
    \boldsymbol{\Omega}^{\chi}_{\mathbf{k}}= \frac{-\chi}{2} 
\frac{N \hbar v \alpha_N^2 k^{2N-2}_{\bot} }{\epsilon_{\mathbf{k}}^{3}}
\: \left( k_x, k_y, N k_z \right),
\end{equation}}
which clearly exhibits strong anisotropy. In addition to Berry curvature, Bloch electrons possess an intrinsic orbital magnetic moment (OMM) arising from self-rotation of the wave packet. It is given by~\cite{xiao2010berry}:
\begin{align}
\mathbf{m}^{\chi}_\mathbf{k} = 
-\frac{ie}{2\hbar} \text{Im} \left\langle \nabla_{\mathbf{k}} u^{\chi} \left| \left[ \epsilon_0(\mathbf{k}) - H^{\chi}(\mathbf{k}) \right] \right| \nabla_{\mathbf{k}} u^{\chi} \right\rangle.
\end{align}
For two-band systems, this simplifies to a direct relation with the Berry curvature~\cite{xiao2005berry,chang1996berry}:
\begin{align}
\mathbf{m}^{\chi}_\mathbf{k}
= \frac{e}{\hbar} \epsilon_0(\mathbf{k}) \, \boldsymbol{\Omega}^{\chi}_\mathbf{k}.
\label{eq:OMM_DW}
\end{align}

Thus, for the multi-Weyl system, the orbital magnetic moment inherits the strong anisotropy of the Berry curvature and is significantly enhanced near the nodal point due to the quadratic dispersion. In the presence of an external magnetic field $\mathbf{B}$, the band energy is modified due to coupling with the orbital magnetic moment:~$\tilde {\epsilon}^{\chi}(\mathbf{k}) = \epsilon_{0}(\mathbf{k}) - \eta ~\mathbf{m}^{\chi}_\mathbf{k} \cdot \mathbf{B}$, with $\epsilon_{0}(\mathbf{k})$ bare band dispersion defined in Eq.~(\ref{eq_multi2_new}). This correction further introduces an angular dependence to the Fermi surface.
The explicit form of the modified energy dispersion $\tilde {\epsilon}^{\chi}(\mathbf{k})$ is calculated to be:
\begin{align}
\tilde{\epsilon}^{\chi}(\mathbf{k}) &=
\sqrt{\alpha_N^2 \left(k_x^2+k_y^2\right)^N
+\hbar^2 k_z^2 v_f^2} \nonumber \\
&+\eta
\left(
\frac{\alpha_N^2 B \chi e\, k_z\, N^2 v_f \sin\gamma
\left(k_x^2+k_y^2\right)^{N-1}}
{2\left(\alpha_N^2\left(k_x^2+k_y^2\right)^N
+\hbar^2 k_z^2 v_f^2\right)}\right) \nonumber \\
&+ \eta \left(\frac{\alpha_N^2 B \chi e\, k_x\, N v_f \cos\gamma
\left(k_x^2+k_y^2\right)^{N-1}}
{2\left(\alpha_N^2\left(k_x^2+k_y^2\right)^N
+\hbar^2 k_z^2 v_f^2\right)}
\right).
\end{align}
\begin{figure}
\centering
\includegraphics[width=1\columnwidth]{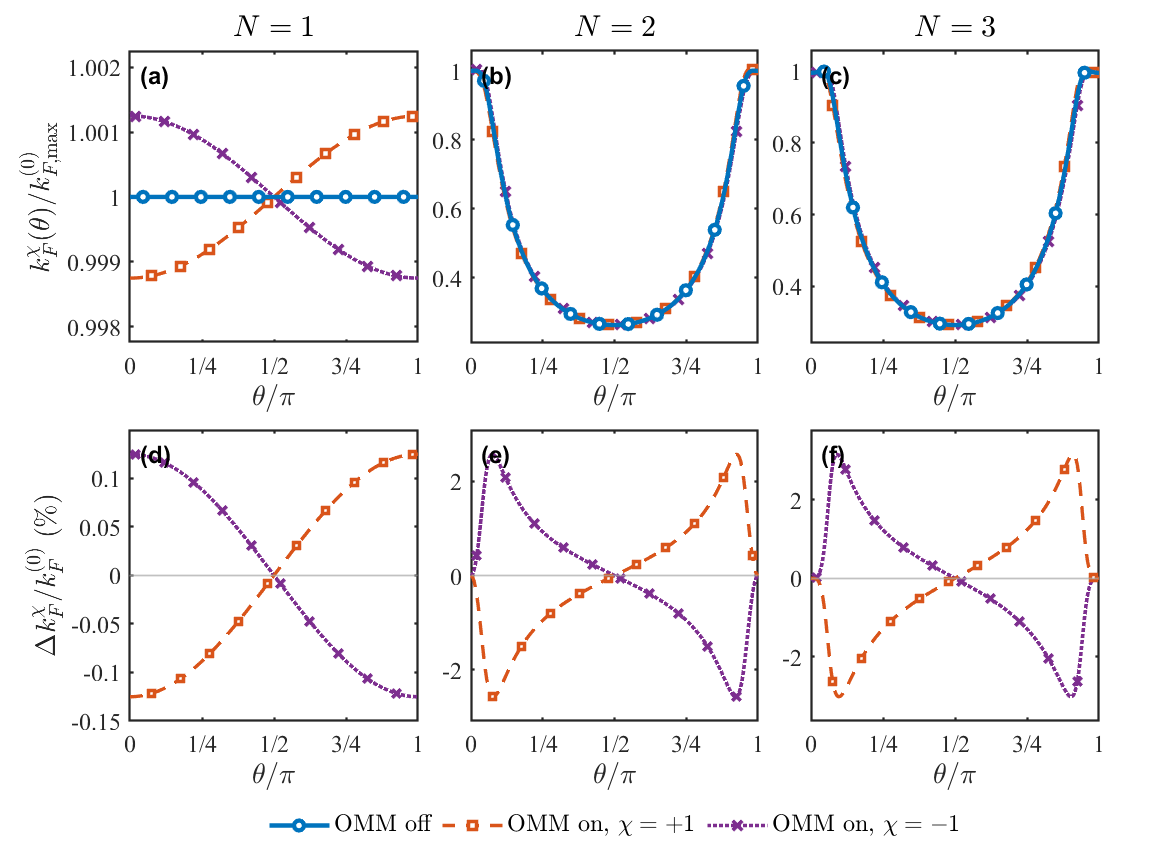}
\caption{Angular dependence of the Fermi momentum $k_F^\chi(\theta)$ for multi-Weyl semimetals with monopole charges (a,d) $N=1$, (b,e) $N=2$, and (c,f) $N=3$, at fixed $\phi=\pi/4$, $B=1$~T, and $\gamma=\pi/2$. The upper panels (a)-(c) show the Fermi momentum normalized by $k_{F,\max}^{(0)}=\max_\theta k_F^{(0)}(\theta)$, where $k_F^{(0)}$ denotes the OMM-free Fermi momentum. The blue solid curves with circles represent the OMM-free case, while the orange dashed curves with squares and purple dotted curves with crosses correspond to $\chi=+1$ and $\chi=-1$, respectively, with OMM included. The lower panels (d)-(f) show the corresponding relative OMM-induced correction, $\Delta k_F^\chi/k_F^{(0)}=[k_F^\chi-k_F^{(0)}]/k_F^{(0)}$, expressed as a percentage. The results illustrate the pronounced Fermi-surface anisotropy for higher monopole charges and the chirality-dependent modifications induced by the OMM.}
\label{fig:fermi_momentum}
\label{Fig:Lambda_n132_omm0nd1.png}
\end{figure}
The modified Fermi surface can be obtained by inverting modified energy dispersion, which leads to a highly anisotropic Fermi contour $k_F^{\chi}(\theta,\phi)$;~$\widetilde\epsilon^\chi\bigl(k_F^\chi(\theta,\phi),\theta,\phi\bigr)=\epsilon_F$ ~\cite{ahmad2021longitudinal,sharma2020sign,varma2026chiral,ahmad2026semiclassical}. Due to the complicated nature of $k_F^{\chi}(\theta,\phi)$, we have calculated it numerically. The well-behaved physical numerical solution used in the present study is plotted in Fig.~\ref{Fig:Lambda_n132_omm0nd1.png}. 

{\color{black}Please} note that, to systematically analyze the role of the orbital magnetic moment, we introduce a control parameter $\eta \in \{0,1\}$. Here, $\eta=0$ corresponds to neglecting OMM effects, while $\eta=1$ includes them fully. This allows us to isolate and quantify the impact of orbital magnetic moment on magnetotransport properties such as longitudinal magnetoconductivity and PHC. {\color{black}The interplay of anisotropic dispersion, enhanced Berry curvature (with monopole charge $\pm N $), and orbital magnetic moment makes multi-Weyl semimetals a fertile platform for exploring nonlinear and anisotropic magnetotransport phenomena.}

The quasiparticle velocity in the presence of the orbital OMM is given by
\begin{align}
\tilde{v}_i^\chi = v_i^\chi 
- \frac{B}{\hbar}\left[
\cos\gamma\,\partial_{k_i} m_x^\chi 
+ \sin\gamma\,\partial_{k_i} m_z^\chi
\right],
\end{align}
where $i=x,y,z$. Explicit expressions are relegated to the appendix ~\ref{Sec.:velocity_components}.
\section{Quasiclassical transport theory}
\label{sec:transport_DW}
The semiclassical dynamics of Bloch electrons in Weyl semimetals, in the presence of external electric ($\mathbf{E}$) and magnetic ($\mathbf{B}$) fields, are governed by the modified equations of motion incorporating Berry curvature effects~\cite{son2012berry,sundaram1999wave,xiao2010berry}:
\begin{align}
\dot{\mathbf{r}}^\chi &= \mathcal{D}^\chi_\mathbf{k} \left[ \mathbf{v}_\mathbf{k}^\chi + \frac{e}{\hbar} \mathbf{E} \times \boldsymbol{\Omega}^\chi_\mathbf{k} + \frac{e}{\hbar} \left( \mathbf{v}_\mathbf{k}^\chi \cdot \boldsymbol{\Omega}^\chi_\mathbf{k} \right) \mathbf{B} \right], \nonumber \\
\dot{\mathbf{k}}^\chi &= -\frac{e}{\hbar} \mathcal{D}^\chi_\mathbf{k} \left[ \mathbf{E} + \mathbf{v}_\mathbf{k}^\chi \times \mathbf{B} + \frac{e}{\hbar} (\mathbf{E} \cdot \mathbf{B}) \boldsymbol{\Omega}^\chi_\mathbf{k} \right],
\label{eq:EOM_DW}
\end{align}
\begin{table*}[t]
\caption{Sign factors for scalar and magnetic point-like impurity channels in Eq.~\eqref{eq:overlap_DW}. 
The indices $1,2,3,4$ correspond to $i=0,x,y,z$, respectively.}
\label{tab:app-signs}
\centering\renewcommand{\arraystretch}{1.15}
\begin{tabular*}{\textwidth}{|c@{\extracolsep{\fill}}cccc|}
\hline\hline
$i$&$\sigma_i$&$C_{i,x}$&$C_{i,y}$&$C_{i,z}$\\\hline
$0$&$I_{2\times2}$&$+1$&$+1$&$+1$\\
$x$&$\sigma_x$&$+1$&$-1$&$-1$\\
$y$&$\sigma_y$&$-1$&$+1$&$-1$\\
$z$&$\sigma_z$&$-1$&$-1$&$+1$\\\hline\hline
\end{tabular*}
\end{table*}
where the phase-space factor is given by: $\mathcal{D}^\chi_\mathbf{k} = \left( 1 + \frac{e}{\hbar} \mathbf{B} \cdot \boldsymbol{\Omega}^\chi_\mathbf{k} \right)^{-1}$~\cite{duval2006comment,xiao2010berry}. The band velocity is obtained from the modified dispersion including orbital OMM: $\tilde{\epsilon}^\chi(\mathbf{k}) = \epsilon_{0}(\mathbf{k}) - \eta \, \mathbf{m}^\chi_\mathbf{k} \cdot \mathbf{B}$, leading to: $\mathbf{v}_\mathbf{k}^\chi = \frac{1}{\hbar} \nabla_\mathbf{k} \tilde{\epsilon}^\chi(\mathbf{k})$. We emphasize that for multi Weyl fermions, both $\boldsymbol{\Omega}_\mathbf{k}$ and $\mathbf{m}_\mathbf{k}$ are strongly anisotropic functions of momentum, which plays a crucial role in determining magnetotransport.

To study transport, we employ the semiclassical Boltzmann equation~\cite{kim2014boltzmann,knoll2020negative,imran2018berry,zyuzin2017magnetotransport}:
\begin{align}
\frac{\partial f^{\chi}_{\mathbf{k}}}{\partial t} + \dot{\mathbf{r}}^{\chi} \cdot \nabla_\mathbf{r} f^{\chi}_{\mathbf{k}} + \dot{\mathbf{k}}^{\chi} \cdot \nabla_\mathbf{k} f^{\chi}_{\mathbf{k}} = I_{\mathrm{coll}}[f^{\chi}_{\mathbf{k}}],
\label{eq:BTE_DW}
\end{align}
where $f^{\chi}_{\mathbf{k}} = f_0 + g^{\chi}_{\mathbf{k}}$. Assuming spatial homogeneity and linear response in $\mathbf{E}$, the deviation function up-to first order in electric field is written as~\cite{morimoto2016semiclassical,sodemann2015quantum,das2022nonlinear,mandal2022chiral,das2022nonlinear,gao2022suppression}:
\begin{align}
g^{\chi}_{\mathbf{k}} = -e \left( \frac{\partial f_0}{\partial \epsilon} \right) \Lambda^{\chi}(\theta,\phi) E,
\label{eq:g1}
\end{align}
with $\mathbf{E} = E \hat{z}$ and $\mathbf{B} = B(\cos\gamma,0,\sin\gamma)$. To calculate $g^{\chi}_{\mathbf{k}}$, we can write the Boltzmann transport equation by using Eqs.~\eqref{eq:EOM_DW} and \eqref{eq:BTE_DW}, and keeping terms up-to-linear order in the electric field as~\cite{bruus2004many,ziman1979principles,sharma2020sign,knoll2020negative,son2013chiral,sodemann2015quantum},
\begin{align}
& -e{\color{black}\mathcal{D}^\chi_{\bf{k}}} \left(\mathbf{E} + \frac{e\mathbf{E} \cdot\mathbf{B}}{\hbar} \boldsymbol{\Omega}_\mathbf{k}^\chi \right)\cdot \mathbf{v}^{\chi}_\mathbf{k}\frac{\partial f_0^\chi}{\partial \epsilon^\chi_\mathbf{k}}\\ \nonumber&= \textcolor{black}{\sum\limits_{\chi',\mathbf{k}'} W^{\chi\chi'}_{\mathbf{k}\mathbf{k}'} (g^{\chi'}_{\mathbf{k}'} - g^{\chi}_\mathbf{k})},
\label{Eq:Eq_boltz_g11}
\end{align}
In low temperature limit ($T\to 0$), using $\sum_{\textbf{k}',\chi'} \to \mathcal{V}\int (\mathcal{D}^{\chi'}_{\textbf{k}})^{-1} d^3 k'/(2\pi)^3 \equiv \mathcal{V}/(2\pi)^3\iiint (\mathcal{D}^{\chi'}_{\textbf{k}'})^{-1} k'^2 \sin \theta' d\theta' d \phi' dk'  $ and exploiting the property of $\frac{\partial f_0^\chi}{\partial \epsilon^\chi_\mathbf{k}}$, above equation can be simplified over the Fermi surface {$\epsilon^{\chi}_{\mathbf{k}}=\epsilon^{\chi'}_{\mathbf{k}'}=\epsilon_F$} as,
\begin{align}
 &  h^{\chi}_{z,\bf{k}}(\theta,\phi) + \frac{\Lambda^{\chi}_{z,\bf{k}}(\theta,\phi)}{{\tau}^{\chi}_{\bf{k}}(\theta,\phi)} \\ \nonumber
 &=\sum_{\chi'} \Pi^{\chi\chi'}\iint \Bigg[\frac{(k'^{\chi'})^3 \sin\theta'}{|\mathbf{v}^{\chi'}_{\mathbf{k}'}\cdot \mathbf{k}'^{\chi'}|} (\mathcal{D}^{\chi'}_{\mathbf{k}'})^{-1}  \Lambda_{z,\bf{k}'}^{\chi'}\Bigg]_{k'=k_F} \nonumber \\
 & \times  \mathcal{T}^{\chi \chi'}   (\theta, \phi, \theta', \phi')  d\theta' d\phi' \nonumber. 
 \label{Eq.:BTE_theta_phi}
\end{align}
Here, $\Pi^{\chi \chi'} = n/ 8\pi^3 \hbar^2$, $h^{\chi}_{z,\bf{k}}(\theta,\phi)=\mathcal{D}^{\chi}_{\mathbf{k}}[v^{\chi}_{z,\textbf{k}}+\frac{eB_z}{\hbar}(\mathbf{\Omega}^{\chi}_{k}\cdot \mathbf{v}^{\chi}_{\mathbf{k}})]$ with $B_z=B\sin{\gamma}$. $\mathcal{T}^{\chi \chi'}$ represents the overlap of the wave function. The collision integral incorporates both intra- and inter-node scattering~\cite{mahan20089,bruus2004many,ziman1979principles,knoll2020negative,ahmad2024geometry}:
\begin{align}
I_{\mathrm{coll}} = \sum_{\chi',\mathbf{k}'} W^{\chi\chi'}_{\mathbf{k}\mathbf{k}'} \left( f^{\chi'}_{\mathbf{k}'} - f^{\chi}_{\mathbf{k}} \right),
\end{align}
where the scattering rate is obtained from Fermi's golden rule~\cite{mahan20089,bruus2004many,ziman1979principles,knoll2020negative,ahmad2024geometry}: $W^{\chi\chi'}_{\mathbf{k}\mathbf{k}'} = \frac{2\pi n}{\hbar \mathcal{V}}  
\left| \langle u^{\chi'}(\mathbf{k}') |U^{\chi\chi'}_{\mathbf{k}\mathbf{k}'}| u^{\chi}(\mathbf{k}) \rangle \right|^2
\delta(\epsilon_{\mathbf{k}'} - \epsilon_{\mathbf{k}})$. Here, \(n\) is the impurity concentration, \(\mathcal{V}\) is the system volume, \(U^{\chi\chi'}_{\mathbf{k}\mathbf{k}'}\) is the scattering potential profile. In general, scattering center can control the chirality of the wave packet after scattering and therefore we may write: $n =n^{\chi \chi'}$. in the current study, for simplicity and mathematical convenience, we keep $n$ chirality index free. Within the present formalism, one can study non-magnetic point-like impurities along with magnetic impurities, therefor choice of \(U^{\chi\chi'}_{\mathbf{k}\mathbf{k}'}\) is generalized to incorporate both magnetic and non-magnetic point-like scattering centers~\cite{varma2024magnetotransport,varma2026chiral}. In general, is written as
\begin{equation}
U^{\chi\chi'}_{\mathbf{k}\mathbf{k}'} = U^{\chi\chi'} \sigma_i,
\end{equation}
where \(i = 0, x, y, z\), and \(U^{\chi\chi'}\) characterizes the internode and intranode scattering processes. It is important to note that, within our framework, the relative contribution of these two scattering channels can be controlled through the dimensionless parameter defined as
\begin{equation}
\alpha = \frac{U^{\chi\ne\chi'}}{U^{\chi\chi}}.
\end{equation}
For multi-Weyl fermions, the eigen-states possess an \(N\phi\) winding, resulting in an overlap structure that is qualitatively different from that of linear Weyl systems~\cite{ahmad2021longitudinal,ahmad2026magnetotransport,ahmad2023longitudinal,ahmad2024geometry,ahmad2025chiral,ahmad2025longitudinal,ahmad2025strain,ahmad2026magnetotransport,varma2026strain}. The overlap factor evaluates to:

\begin{align}
&\mathcal{T}^{\chi\chi'}_{i}(\theta,\theta',\phi,\phi') =  \frac{|U^{\chi\chi'}|^2}{2}[1 +\chi\chi' \{C_{i,z} ~u_z^\chi(\theta,\phi) ~u_z^{\chi'}(\theta',\phi')\nonumber \\  &+ C_{i,x} ~u_x^\chi(\theta,\phi) ~u_x^{\chi'}(\theta',\phi') + C_{i,y} ~u_y^\chi(\theta,\phi) ~u_y^{\chi'}(\theta',\phi') \}].
\label{eq:overlap_DW}
\end{align}
Please note that, we have use following notation, 
\begin{align}
u_z^\chi&=\frac{\beta^\chi\cos\theta}{R^\chi},
u_x^\chi=\frac{\sin^N\theta\cos(N\phi)}{R^\chi},\nonumber \\
u_y^\chi&=\frac{\sin^N\theta\sin(N\phi)}{R^\chi}, \nonumber \\
\beta^\chi(\theta,\phi)&=\frac{\hbar v_F}{\alpha_N[k_F^\chi(\theta,\phi)]^{N-1}},\nonumber \\  R^\chi(\theta,\phi)&=\sqrt{[\beta^\chi(\theta,\phi)]^2\cos^2\theta+\sin^{2N}\theta}.
\label{eq:app-u}
\end{align}
Thus $(u_x^\chi)^2+(u_y^\chi)^2+(u_z^\chi)^2=1$.
The appearance of $\cos N(\phi-\phi')$ is a direct consequence of the multi topological charge and leads to enhanced angular anisotropy in scattering. In general, the dimensionless parameter, $\beta^{\chi}$, is a function of $\theta $ and $\phi$ through the constant Fermi contour $k_F^{\chi}(\theta,\phi)$: $\beta^{\chi} (\theta,\phi) = \frac{\hbar v_F}{\alpha_N (k_F^{\chi})^{N-1}}$. The momentum-dependent scattering rate is then defined as: $\frac{1}{\tau^{\chi}(\theta,\phi)} 
= \sum_{\chi'} \int \frac{d^3\mathbf{k}'}{(2\pi)^3}
(\mathcal{D}^{\chi'}_{\mathbf{k}'})^{-1}
W^{\chi\chi'}_{\mathbf{k}\mathbf{k}'}$. Using the energy-conserving delta function, the radial integration can be performed, yielding:
\begin{align}
\frac{1}{\tau^{\chi}_{i}(\theta,\phi)} 
= \sum_{\chi'} \Pi^{\chi\chi'} \iint d\theta' d\phi' \,
\mathcal{F}^{\chi'}(\theta',\phi') 
\, \mathcal{T}^{\chi\chi'}_{i}(\theta,\theta',\phi,\phi'),
\label{Eq.:tau_thetaphi}
\end{align}
where $\mathcal{F}^{\chi'}$ encodes the anisotropic density of states: $\mathcal{F}^{\chi}(\theta,\phi) =
\Bigg[\frac{k^3~\sin\theta}{|\mathbf{v}^\chi \cdot \mathbf{k}|}\, (\mathcal{D}^\chi_{\mathbf{k}})^{-1}\Bigg]_{k'=k_F}$. In order to solve Eq.~\ref{Eq.:BTE_theta_phi}, we have to choose an ansatz. Motivated by the angular dependencies of the overlap function  $\mathcal{T}^{\chi\chi'}(\theta,\phi)$, one can choose the following ansatz:
\begin{align}
    &\Lambda^{\chi}_{z,\mathbf{k},i} = \textcolor{black}{{\tau}^{\chi}_{\textbf{k}}(\theta,\phi)}\big[d^{\chi}_{i}-h^{\chi}_{z,\mathbf{k}} \\ \nonumber
    + &\frac{a^{\chi}_{i} \beta^{\chi} \cos{\theta} +b^{\chi}_{i}\sin^{N}{\theta}cos{N\phi}+c^{\chi}_{i}\sin^{N}{\theta}\sin{N\phi}} {\sqrt{[\beta^{\chi}]^2 \cos^2\theta + \sin^{2N}\theta} }\big],\nonumber\\
    &\equiv {\tau}^{\chi}_{\textbf{k}}(\theta,\phi)\big[d^{\chi}_{i}-h^{\chi}_{z,\mathbf{k}} +a^{\chi}_{i} u_z^\chi(\theta,\phi)\nonumber + b^{\chi}_{i} u_x^\chi(\theta,\phi)  ,\nonumber\\
    & + c^{\chi}_{i} u_y^\chi(\theta,\phi) \big]
\end{align}
where, $a^\chi_{i}$, $b^\chi_{i}$, $c^{\chi}_{i}$ and  $d^\chi_{i}$ are the unknown coefficients for $i\in(0,1,2,3)$, having the dimension of velocity. The above Eq. (\ref{Eq.:BTE_theta_phi}) can be written in following form
\begin{align}
&d^{\chi}_{i}+\frac{a^{\chi}_{i} \beta^{\chi} \cos{\theta} +b^{\chi}_{i}\sin^{N}{\theta}cos{N\phi}+c^{\chi}_{i}\sin^{N}{\theta}\sin{N\phi}} {\sqrt{[\beta^{\chi}]^2 \cos^2\theta + \sin^{2N}\theta} } \nonumber \\
&=\sum_{\chi'}\Pi^{\chi\chi'}\iint \Bigg[\frac{(k'^{\chi'})^3}{|\mathbf{v}^{\chi'}_{{\bf k}'}\cdot{\mathbf{k}^{\chi'}}|} \widetilde{\tau}^{\chi'}_{\textbf{k}'}(\theta,\phi) (\mathcal{D}^{\chi'}_{\mathbf{k}'})^{-1}\Bigg]_{k'=k_F}   \nonumber \\
&\times\bigg[ d^{\chi'}_{i}-\big[h^{\chi'}_{z,\textbf{k}'}\big]_{k'=k_F}+ \nonumber \\
&\times\frac{a^{\chi'}_{i} \beta^{\chi'} \cos{\theta'} +b^{\chi'}_{i} \sin^{N}{\theta'}cos{N\phi'}+c^{\chi'}_{i} \sin^{N}{\theta'}\sin{N\phi'}} {\sqrt{[\beta^{\chi'}]^2 \cos^2\theta' + \sin^{2N}\theta'}} \bigg] \nonumber \\
&\times
 \mathcal{T}^{\chi\chi'}_{i}(\theta, \phi,\theta', \phi') ~\sin{\theta'}d\theta'd\phi'.
\label{Boltzman_final}
\end{align}

After equating the coefficients associated with $(1, \beta^{\chi}\cos\theta/\sqrt{[\beta^{\chi}]^2 \cos^2\theta + \sin^{2N}\theta},\\ ~\sin^{N}\theta \cos N \phi/\sqrt{[\beta^{\chi}]^2 \cos^2\theta + \sin^{2N}\theta},\\~ \sin^{N}\theta \sin N \phi/\sqrt{[\beta^{\chi}]^2 \cos^2\theta + \sin^{2N}\theta})$ terms from both sides of the above Eq. (\ref{Boltzman_final}), one can solve the coefficients using a matrix form. Once this set of equations is expanded for $\chi=\pm 1$ and for each impurity type, it actually yields seven linearly independent equations consisting of eight variables that must be solved.  The particle number conservation provides an additional restriction. The particle number conservation is given by $\sum\limits_{\chi}\sum\limits_{\mathbf{k}} g^{\chi}_\mathbf{k}= 0$ which takes the following form 
\begin{align}
&\sum_{\chi}\iint  \Bigg[\frac{(k^\chi)^3{\tau}^{\chi}_{\bf{k}}(\theta,\phi)}{|\mathbf{v}^{\chi}_{{\bf k}}\cdot{\mathbf{k}^{\chi}}|} \Bigg]_{k=k_F} ~ 
\bigg[d^{\chi}_{i}-\big[h^{\chi}_{z,\textbf{k}}\big]_{k=k_F}\nonumber \\
&
+\frac{a^{\chi}_{i} \beta^{\chi} \cos{\theta} +b^{\chi}_{i}\sin^{N}{\theta}cos{N\phi}+c^{\chi}_{i}\sin^{N}{\theta}\sin{N\phi}} {\sqrt{[\beta^{\chi}]^2 \cos^2\theta + \sin^{2N}\theta} }\bigg]\nonumber \\
&~~~~~~~~~~~~~~~~~~~~~~~~~~~~~~~~~~~~~
\times \sin{\theta} ~d\theta d\phi= 0.
\label{Eq_sumgk}
\end{align}
For eight unknown coefficients ($d^{\pm}_{i}, a^{\pm }_{i}, b^{\pm }_{i}, c^{\pm }_{i}$), Eqs. (\ref{Boltzman_final}) and (\ref{Eq_sumgk}) are simultaneously solved with Eq.~(\ref{Eq.:tau_thetaphi}). Due to the intricate structure of equations, all two-dimensional integrals with respect to $\theta'$ and $\phi'$, the solutions of the coupled \textcolor{black}{equations are carried out numerically}. 

Finally, the current is computed as:
\begin{align}
\mathbf{J} = -e \sum_{\chi,\mathbf{k}} f^\chi_\mathbf{k} \dot{\mathbf{r}}^\chi,
\end{align}

As a consistency check of the present Boltzmann formalism,
we examine the analytically solvable zero-field limit
for a conventional single-Weyl dispersion ($N=1$).
In this limit, the two Weyl nodes have identical isotropic
Fermi surfaces, and the momentum-dependent collision
integral reduces to a set of coupled equations that can
be solved analytically. The resulting longitudinal
conductivity is positive and exhibits distinct
impurity-dependent transport relaxation times, with
$\sigma_{zz}^{(0)}(0)=\sigma_{zz}^{(z)}(0)$ and
$\sigma_{zz}^{(x)}(0)=\sigma_{zz}^{(y)}(0)$.
The detailed derivation, including the analytical
transport relaxation times and zero-field conductivity
for all four impurity channels, is presented in
Appendix~\ref{app:zero_field_single_weyl}. \textcolor{black}{ Unless stated otherwise, the numerical calculations are performed at a Fermi energy $\epsilon_F=50\,\mathrm{meV}$, with $v_f=1.0\times10^{5}\,\mathrm{m\,s^{-1}}$ and $m=2.75\times10^{-32}\,\mathrm{kg}$. The dispersion coefficients are chosen as $\alpha_1=\hbar v_f$, $\alpha_2=\hbar^2/(2m)$, and $\alpha_3=mv_f^2/\lambda^3$, where $\lambda=k_F/10.5$ and $k_F=\epsilon_F/(\hbar v_f)$ denotes the longitudinal Fermi wave vector at zero magnetic field. Here, $\lambda$ parametrizes the cubic dispersion coefficient and is not imposed as a momentum cutoff. These parameters define a phenomenological continuum model rather than a fit to a specific material. We consider monopole charges $N=1,2,3$ and use identical intravalley scattering strengths for the two nodes, while varying the relative intervalley scattering strength $\alpha$. The magnetic field lies in the $xz$ plane, $\mathbf{B}=B(\cos\gamma,0,\sin\gamma)$, with the electric field directed along $\hat{\mathbf{z}}$. The orbital magnetic moment is included unless explicitly stated otherwise.}

\section{Results and discussion}
\label{Sec.:Results and discussion}

\subsection{LMC without OMM in multi-WSMs}
\label{Sec.:LMC without OMM in multi-WSMs}
We begin by analyzing the LMC in the absence of the OMM for multi-Weyl semimetals with topological charge $N=1$, $2$, $3$, for magnetic and non-magnetic impurity sites. For all three values of $N$, the normalized LMC, $\delta\sigma_{zz}(B)/\sigma_{zz}(0)$, remains positive and quadratic dependence on the magnetic field. The response is minimum at $B=0$ and increases monotonically with increasing field strength and decreases with increasing impurity-induced intervalley scattering strength $\alpha$, which is the characteristic signature of chiral-anomaly-assisted longitudinal transport in Weyl systems ~\cite{knoll2020negative,sharma2020sign,ahmad2021longitudinal,das2019linear}. This benchmarks our numerical results and sets the stage to study the LMC in multi-WSMs. We have not included it explicitly. The persistence of a positive $B^{2}$ dependence for $N>1$ is a nontrivial property of multi-Weyl semimetals. Although higher-order Weyl nodes possess strongly anisotropic nonlinear dispersion in the transverse plane, the anomaly-induced component of the LMC remains quadratic in the weak-field semiclassical regime and is largely insensitive to the field direction \cite{dantas2018magnetotransport}. This behavior originates from Berry-curvature-modified semiclassical dynamics, where parallel electric and magnetic fields generate a nonequilibrium imbalance between Weyl nodes of opposite chirality through the anomaly term proportional to $\mathbf{B}(\mathbf{E}\cdot\mathbf{B})$. The resulting axial charge accumulation gives rise to an additional current along the magnetic-field direction, producing the positive longitudinal magnetoconductivity originally identified in Weyl semimetals and subsequently generalized to multi-Weyl systems \cite{son2013chiral,dantas2018magnetotransport}.
The zero-field conductivity for $N=1$ also provides
an analytical benchmark for the impurity-dependent
collision integral, as discussed in
Appendix~\ref{app:zero_field_single_weyl}. It confirms that the conductivity strongly depends on intervalley scattering for the $\sigma_0$ and $\sigma_z$ impurities compared with the $\sigma_x$ and $\sigma_y$ impurities.\\

\begin{figure*}
\centering
\includegraphics[width=1.95\columnwidth]{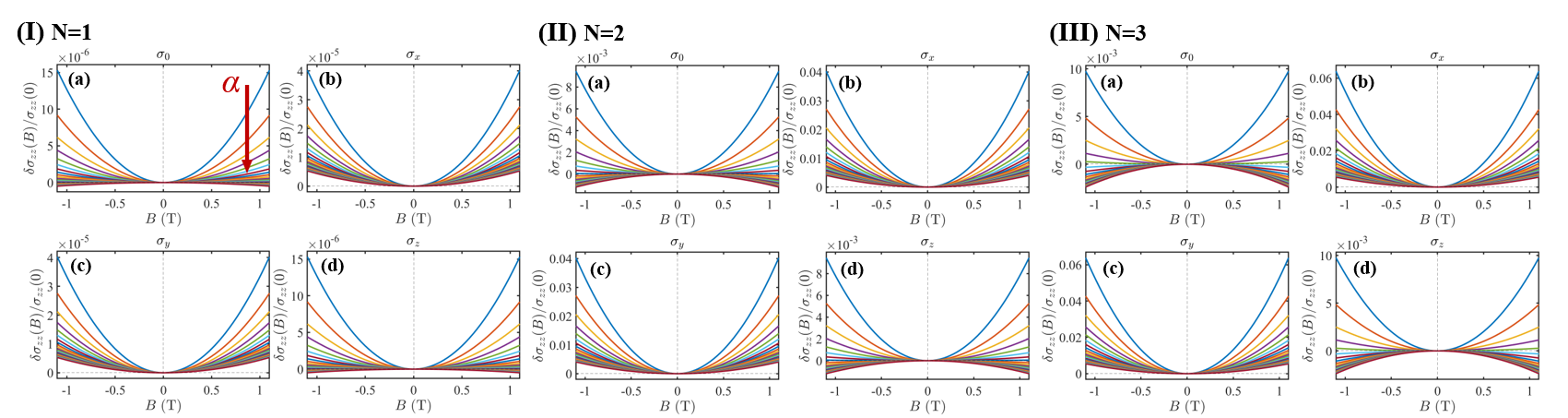}
\caption{Normalized longitudinal magnetoconductivity $\delta\sigma_{zz}(B)/\sigma_{zz}(0)$, where $\delta\sigma_{zz}(B)=\sigma_{zz}(B)-\sigma_{zz}(0)$, as a function of magnetic field $B$ for different intervalley scattering strengths $\alpha$, including the orbital magnetic moment (OMM), with $\mathbf{E}\parallel\mathbf{B}\parallel\hat{\mathbf{z}}$. The left, middle, and right groups correspond to monopole charges $N=1,2,3$, respectively. Within each group, panels (a)--(d) represent impurity potentials $\sigma_0=I$, $\sigma_x$, $\sigma_y$, and $\sigma_z$, respectively. Different colored curves correspond to different values of $\alpha$, with the arrow indicating increasing $\alpha$. The response is approximately quadratic in $B$. Increasing intervalley scattering drives a reversal from positive to negative LMC for $\sigma_0$ and $\sigma_z$, whereas the LMC remains positive for $\sigma_x$ and $\sigma_y$ over the displayed parameter range. The responses coincide for the pairs $(\sigma_0,\sigma_z)$ and $(\sigma_x,\sigma_y)$. Dashed reference lines indicate $B=0$ and zero magnetoconductivity.}
\label{Fig:N123_LMC_vs_B_alp_vary_OMM1}
\end{figure*}
\begin{figure*}
\centering
\includegraphics[width=1.95\columnwidth]{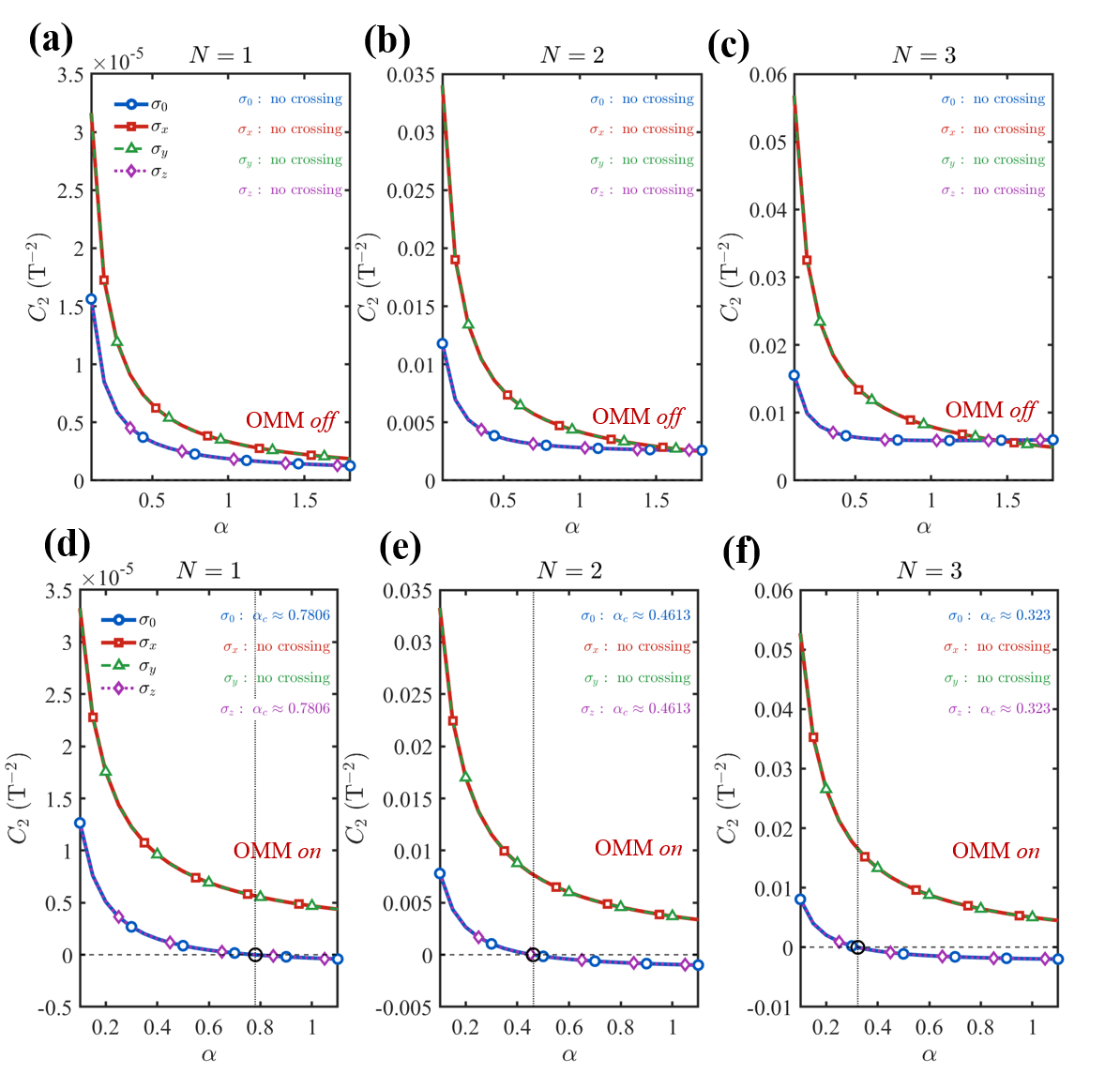}
\caption{Quadratic coefficient $C_2$ of the normalized longitudinal magnetoconductivity, $\delta\sigma_{zz}(B)/\sigma_{zz}(0)=C_0+C_1B+C_2B^2$, as a function of intervalley scattering strength $\alpha$ for $\mathbf{E}\parallel\mathbf{B}\parallel\hat{\mathbf{z}}$. Panels (a)--(c) show the results without the orbital magnetic moment (OMM), while panels (d)--(f) include the OMM, for monopole charges $N=1,2,3$, respectively. Different colors, line styles, and markers distinguish the impurity potentials $\sigma_0=I$, $\sigma_x$, $\sigma_y$, and $\sigma_z$; the $\sigma_0$ and $\sigma_z$ curves overlap, as do the $\sigma_x$ and $\sigma_y$ curves. Without OMM, $C_2$ remains positive throughout the displayed range. Including OMM produces a sign reversal for $\sigma_0$ and $\sigma_z$, with critical scattering strengths $\alpha_c\approx0.7806$, $0.4613$, and $0.323$ for $N=1,2,3$, respectively, whereas no sign reversal is observed for $\sigma_x$ and $\sigma_y$ within the displayed range. Horizontal dashed lines indicate $C_2=0$, and vertical dotted lines and black circles mark the interpolated sign-reversal points.}
\label{Fig:LMC_vs_alp_OMM01}
\end{figure*}
A striking feature we observe is the enhancement of the LMC with increasing topological charge $N$, which scales differently from previously found power laws, which we have explained below. It is expected that as the monopole charge increases from $N=1$ to $N=3$, the magnitude of the LMC grows, while its quadratic field dependence is preserved. This enhancement is a direct consequence of the larger Berry-flux monopole charge associated with higher-order Weyl nodes. The Berry curvature acts as an effective magnetic field in momentum space, and its integrated flux through a closed surface enclosing a Weyl node is quantized by the topological charge. Consequently, the anomaly coefficient itself is proportional to the winding number of the node \cite{huang2017topological,huang2016new}, leading to a stronger chiral charge pumping process and a correspondingly larger longitudinal response. Kinetic-theory analyses of multi-Weyl semimetals further predict that the anomaly-induced LMC scales approximately as $N^{3}$ in the semiclassical regime \cite{dantas2018magnetotransport,nag2021magneto}. A few works also suggest some deviation from the $N^3$ dependence~\cite{medel2024electric}. In our numerical analysis, we found that the power depends upon the actual position of the Fermi surface with respect to the Weyl node. We will come later to the dependence of the conductivities on topological charge $N$.

The effect of intervalley scattering is captured through the parameter $\alpha$. Increasing $\alpha$ systematically suppresses the LMC for all values of $N$, indicating a reduction of the anomaly-induced contribution~\cite{sharma2020sign,sharma2023decoupling,ahmad2021longitudinal,ahmad2026magnetotransport,ahmad2023longitudinal,ahmad2024geometry,ahmad2025chiral,ahmad2025longitudinal,ahmad2025strain,ahmad2026magnetotransport,varma2026strain,varma2024magnetotransport,varma2026chiral,sharma2017chiral}. In the two-constant-relaxation-time picture, intervalley scattering governs the relaxation of axial charge, whereas intravalley scattering primarily restores the isotropy of the distribution function. A stronger intervalley scattering rate therefore shortens the lifetime of the chiral imbalance generated by the anomaly and reduces the additional current flowing along the magnetic field \cite{dantas2018magnetotransport,sharma2020sign,sharma2023decoupling,ahmad2021longitudinal,son2012chiral}. Importantly, despite this quantitative suppression, the LMC remains positive throughout the entire range of $\alpha$ considered. No sign reversal is observed in the absence of OMM, indicating that intervalley scattering alone does not qualitatively alter the nature of the magnetotransport response within the present parameter regime for isotropic ($N=1$) as well as anisotropic ($N>1$) Fermi surfaces and so for magnetic as well as non-magnetic impurity sites. This confirms that the Fermi surface geometry is unrelated to the sign of LMC. Another notable aspect is that increasing $\alpha$ does not change the symmetry or functional form of the conductivity for any of the three WSM types considered. The response remains an even function of magnetic field, $\sigma_{zz}(B)=\sigma_{zz}(-B)$, and retains its quadratic field dependence for all three topological charges. Thus, within the conventional Berry-curvature-corrected semiclassical framework, intervalley scattering acts only as a damping mechanism for the anomaly-induced conductivity and does not introduce competing contributions capable of driving the system into a negative-LMC regime.

These results establish the OMM-free response as the reference state of the system. The positive quadratic magnetoconductivity and its enhancement with increasing topological charge are fully consistent with the established kinetic theory of multi-Weyl semimetals ~\cite{nag2021magneto,huang2017topological,park2017semiclassical,dantas2018magnetotransport}. The absence of any qualitative transition with increasing $\alpha$ provides a crucial benchmark for the next subsection, where we show that the inclusion of the orbital magnetic moment introduces additional intrinsic geometric contributions that fundamentally modify the magnetotransport response and eventually drive a sign reversal of the LMC.
\subsection{LMC with OMM in multi-WSMs}
\label{Sec.:Longitudinal magnetoconductivity with OMM in multi-WSMs}

We now examine the LMC after incorporating the OMM correction into the semiclassical dynamics, as shown in Fig.~\ref{Fig:N123_LMC_vs_B_alp_vary_OMM1}. In contrast to the OMM-\textit{off} case discussed above, the inclusion of the OMM qualitatively modifies the magnetic-field response. While the conductivity remains an even function of magnetic field and retains quadratic dependence in the weak-field regime, increasing the intervalley scattering strength $\alpha$ now drives the system through a sequence of distinct transport regimes, culminating in a complete sign reversal of the LMC at sufficiently large $\alpha$. This behavior is observed for all three topological charges, although the transition becomes increasingly pronounced with increasing $N$. We note that the qualitative behavior of LMC for $\sigma_0$ and $\sigma_z$ is the same, and so for $\sigma_x$ and $\sigma_y$. This is consistent with the previous studies on magnetotransport in the presence of magnetic impurity sites~\cite{varma2024magnetotransport,varma2026chiral}.
The origin of this qualitative change lies in the fact that the OMM is not merely an additive correction to the conductivity, but enters the semiclassical transport theory through a field-dependent modification of the quasiparticle energy, $\tilde{\varepsilon}_{\mathbf{k}}=\varepsilon_{\mathbf{k}}-\mathbf{m}_{\mathbf{k}}\!\cdot\!\mathbf{B}$, where $\mathbf{m}_{\mathbf{k}}$ is the orbital magnetic moment of the Bloch wave packet. This correction modifies the group velocity, the Berry-curvature-dependent phase-space measure, and the nonequilibrium distribution function simultaneously. In summary, the longitudinal magnetoconductivity receives additional intrinsic geometric contributions that appear at leading first order in the magnetic field and are absent in the conventional first-order semiclassical description where OMM effect disregarded~\cite{gao2017intrinsic,gao2014field,gao2015geometrical,chang2008berry,knoll2020negative}. These geometric contributions originate from the interplay between the OMM and the Berry curvature, which together reconstruct the effective quasiparticle dynamics in a magnetic field~\cite{knoll2020negative,xiao2010berry,xiao2005berry}.

The numerical results reveal a competition between the positive chiral-anomaly-assisted magnetoconductivity and the corrections arising from the interplay of Berry curvature and orbital magnetic moment (OMM). In the weak intervalley scattering regime, the chiral imbalance generated by the parallel electric and magnetic fields relaxes slowly, resulting in a positive quadratic LMC. As the relative intervalley scattering strength $\alpha$ increases, the enhanced relaxation of the chiral imbalance progressively suppresses this positive contribution. In the presence of OMM, the magnetic-field-induced modifications of the band dispersion, carrier velocity, and scattering processes further alter the longitudinal transport response. Importantly, as demonstrated by Knoll \textit{et al.}, the negative LMC cannot generally be attributed to an intrinsically negative OMM-only contribution, but instead emerges from the combined effects of Berry curvature and OMM under sufficiently strong intervalley scattering. Beyond a critical value of $\alpha$, these combined corrections outweigh the positive anomaly-assisted contribution, leading to a negative quadratic LMC. The observed sign reversal therefore reflects the interplay between chiral-anomaly-assisted transport, OMM-induced modifications, and intervalley scattering, consistent with previous semiclassical Boltzmann studies~\cite{knoll2020negative,sharma2020sign}.

This mechanism is consistent with the weak-field semiclassical analysis of Weyl semimetals by Knoll \textit{et al.}, who demonstrated that the combined effects of BC and OMM can produce negative longitudinal magnetoconductivity when intervalley scattering is sufficiently strong~\cite{knoll2020negative}. Importantly, our numerical results establish that a similar sign-reversal mechanism persists in multi-Weyl semimetals and becomes increasingly pronounced with higher monopole charge $N$. As illustrated in Figs.~\ref{Fig:N123_LMC_vs_B_alp_vary_OMM1} and \ref{Fig:LMC_vs_alp_OMM01}, increasing $N$ systematically lowers the critical intervalley scattering strength required for the transition from positive to negative LMC in the $\sigma_0$ and $\sigma_z$ impurity channels. Specifically, the critical values decrease from $\alpha_c \approx 0.7806$ for $N=1$ to $0.4613$ and $0.323$ for $N=2$ and $N=3$, respectively. Furthermore, the magnitude of the negative quadratic magnetoconductivity is substantially enhanced for higher-charge Weyl nodes within the investigated parameter regime. This behavior suggests that the modifications of Berry curvature and OMM associated with higher monopole charge, together with the anisotropic band dispersion and momentum-dependent scattering processes, strengthen the negative corrections to the longitudinal transport response relative to the positive anomaly-assisted contribution. Consequently, higher-charge Weyl nodes exhibit a more pronounced sign reversal at weaker intervalley scattering strengths. In contrast, the LMC remains positive for the $\sigma_x$ and $\sigma_y$ impurity channels throughout the investigated range, demonstrating that the emergence and enhancement of negative LMC depend sensitively on the nature of impurity scattering.

A particularly notable feature of Fig.~\ref{Fig:LMC_vs_alp_OMM01}(d)-(f) is that the magnitude of the negative quadratic LMC becomes substantially enhanced with increasing topological charge for the $\sigma_0$ and $\sigma_z$ impurity channels. This behavior can be understood in terms of the modified band geometry associated with higher-order Weyl nodes. In multi-Weyl semimetals, both the Berry curvature and the orbital magnetic moment depend strongly on the monopole charge and exhibit increasingly anisotropic momentum-space distributions, while the chiral anomaly coefficient is proportional to the monopole charge~\cite{huang2017topological,dantas2018magnetotransport}. Consequently, increasing $N$ modifies the relative importance of the positive anomaly-assisted contribution and the negative corrections arising from the combined effects of Berry curvature and OMM. The pronounced negative LMC observed for the triple-Weyl case suggests that these geometric corrections, together with the momentum-dependent scattering processes, become increasingly effective in overcoming the positive anomaly-assisted response within the investigated parameter regime. Thus, the enhancement of negative magnetoconductivity with increasing $N$ reflects the combined influence of higher monopole charge, anisotropic band geometry, and intervalley scattering, rather than a simple scaling of the OMM contribution alone.

\begin{figure}
\centering
\includegraphics[width=.95\columnwidth]{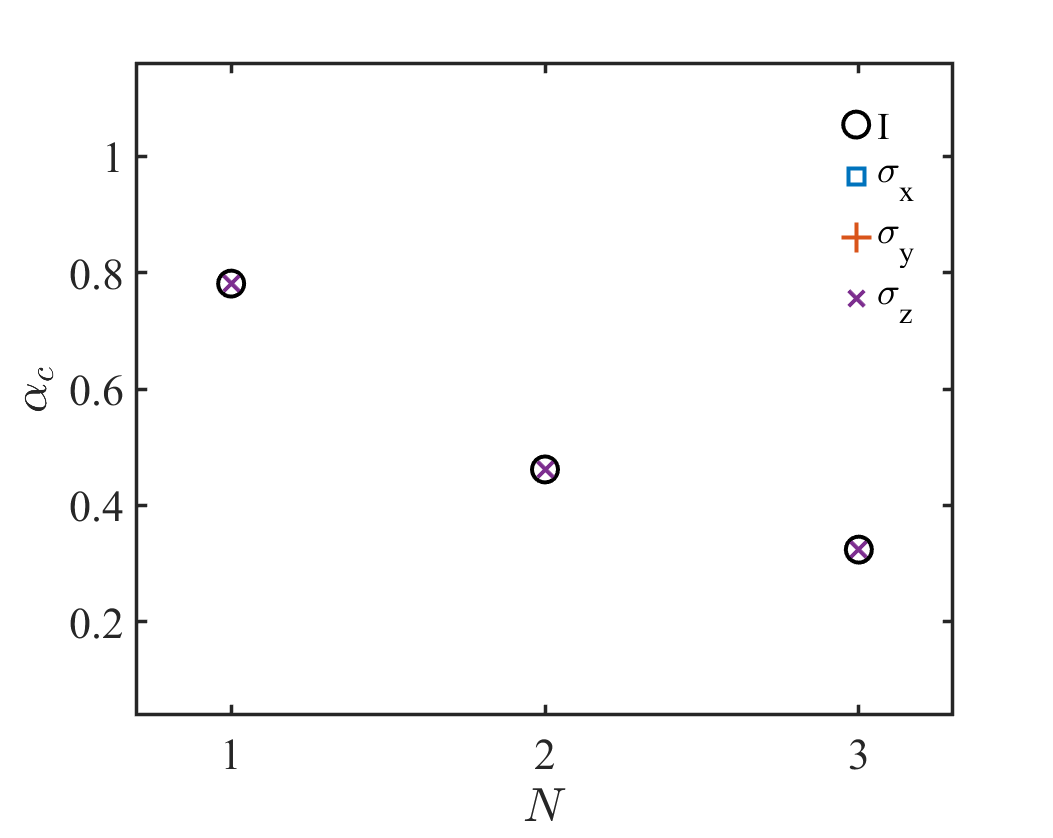}
\caption{Critical intervalley scattering strength
$\alpha_c$ as a function of monopole charge $N$
for the four impurity potentials
$I$, $\sigma_x$, $\sigma_y$, and $\sigma_z$,
with $\mathbf{E}\parallel\mathbf{B}
\parallel\hat{\mathbf{z}}$.
The critical strength is determined from the
sign change of the quadratic coefficient $C_2$,
extracted from
$\delta\sigma_{zz}(B)/\sigma_{zz}(0)
= C_0+C_1B+C_2B^2$.
Values of $\alpha_c$ are estimated by linear
interpolation between adjacent scattering
strengths with opposite signs of $C_2$.
Missing markers indicate that no sign reversal
is resolved within the sampled intervalley
scattering range, while coincident markers
indicate overlapping critical values.}
\label{fig:alpha_c_vs_N}
\end{figure}

To further quantify the influence of monopole charge on the sign reversal of LMC, we examine the critical intervalley scattering strength $\alpha_c$ as a function of $N$, as shown in Fig.~\ref{fig:alpha_c_vs_N}. The critical strength is determined from the condition $C_2(\alpha_c)=0$, where $C_2$ is the quadratic coefficient of the normalized LMC. For the $\sigma_0$ and $\sigma_z$ impurity channels, $\alpha_c$ decreases systematically from approximately $0.7806$ for $N=1$ to $0.4613$ and $0.323$ for $N=2$ and $N=3$, respectively. This behavior demonstrates that higher-charge Weyl nodes require progressively weaker intervalley scattering to undergo the transition from positive to negative LMC. In contrast, no sign reversal is observed for the $\sigma_x$ and $\sigma_y$ impurity channels within the investigated scattering range. The systematic reduction of $\alpha_c$ with increasing $N$ suggests that the combined Berry-curvature- and OMM-related corrections become increasingly effective relative to the positive anomaly-assisted contribution. These findings further emphasize that the sign reversal is governed by the interplay of monopole charge, band geometry, and the nature of impurity scattering.

It is worth emphasizing that the sign reversal observed here cannot be reproduced within the OMM-free semiclassical framework for the model and parameter regime considered. In the absence of OMM, increasing intervalley scattering progressively suppresses the positive quadratic LMC without changing its sign, as demonstrated in Figs.~\ref{Fig:LMC_vs_alp_OMM01}(a)-(c). In contrast, incorporating OMM allows the combined effects of Berry curvature and OMM to produce negative quadratic magnetoconductivity when intervalley scattering is sufficiently strong~\cite{knoll2020negative}. The observed sign reversal therefore highlights the essential role of OMM-dependent geometric corrections in weak-field magnetotransport. Furthermore, the systematic reduction of the critical intervalley scattering strength with increasing monopole charge, as summarized in Fig.~\ref{fig:alpha_c_vs_N}, demonstrates that higher-order Weyl nodes exhibit an enhanced tendency toward negative LMC for the $\sigma_0$ and $\sigma_z$ impurity channels. These results establish the interplay of Berry curvature, OMM, monopole charge, and impurity scattering as a key factor governing the sign and magnitude of longitudinal magnetoconductivity in multi-Weyl semimetals.
\begin{figure*}
\centering
\includegraphics[width=2\columnwidth]{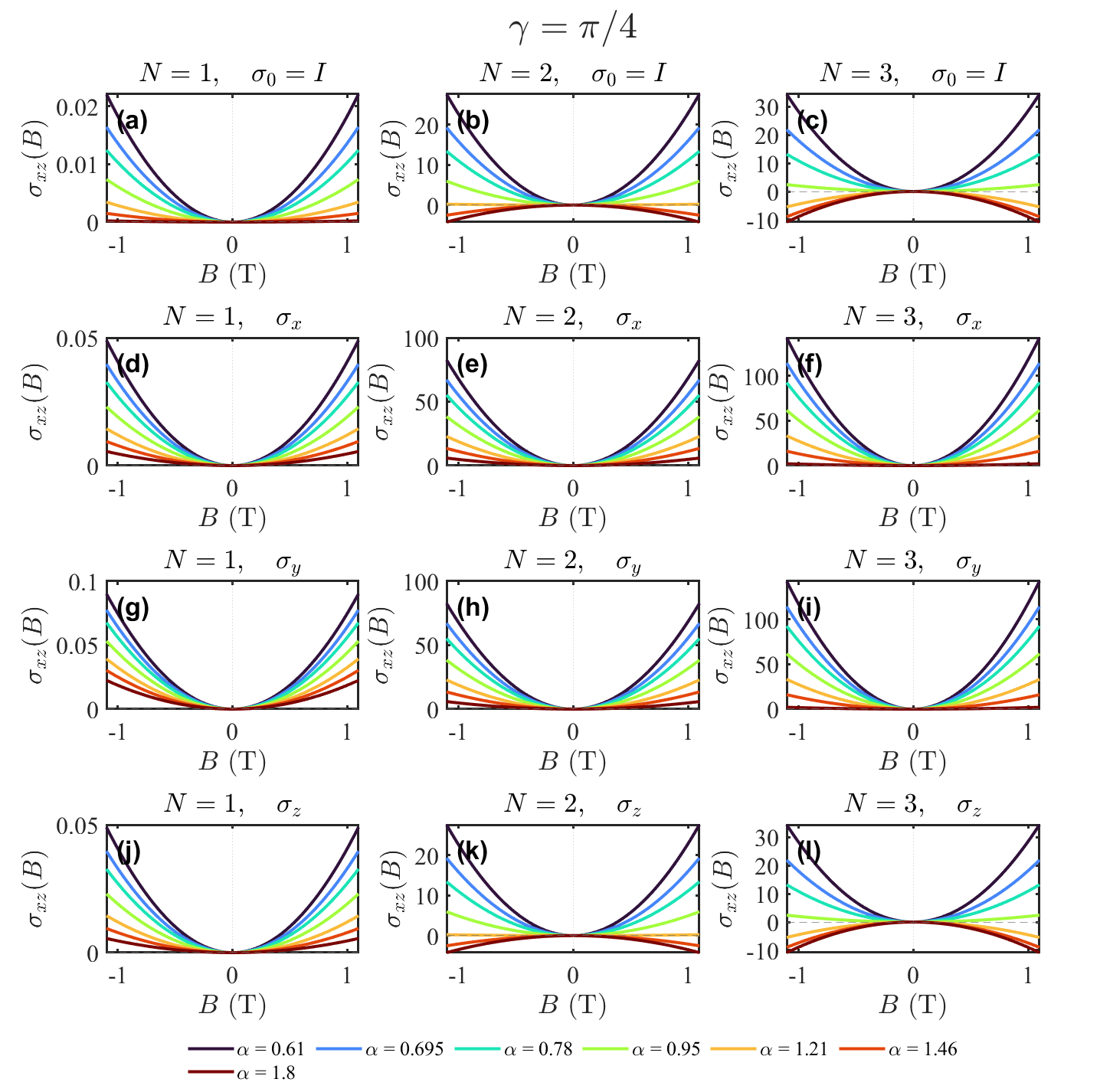}
\caption{PHC, $\sigma_{zx}(B)$, as a function of magnetic field $B$ for different values of the intervalley scattering parameter $\alpha$ in multi-WSMs with topological charge (a) $N=1$, (b) $N=2$, and (c) $N=3$. For $N=1$ and $N=2$, increasing $\alpha$ progressively suppresses the positive PHC, while the overall quadratic magnetic-field dependence and positive sign remain unchanged. Increasing the topological charge enhances the magnitude of the PHC without altering its qualitative behavior. In contrast, for $N=3$, the PHC undergoes a sign reversal in the strong intervalley scattering regime, becoming negative at sufficiently large values of $\alpha$.}
\label{Fig:N123_PHC_vs_B_alp_vary_OMM1.png}
\end{figure*}
\subsection{Magnetic-field dependence of PHC}

We next examine the magnetic-field dependence of the PHC with
OMM included, for a fixed field orientation $\gamma=\pi/4$.
This orientation corresponds to the maximum magnitude of the
leading $\sin 2\gamma$ angular dependence of the planar Hall
response. The corresponding results are shown in
Fig.~\ref{Fig:N123_PHC_vs_B_alp_vary_OMM1.png} for multi-Weyl
semimetals with topological charges $N=1$, $2$, and $3$, for
different intervalley scattering strengths $\alpha$ and impurity
channels. As in the longitudinal response discussed above, the
PHC remains an even function of magnetic field and exhibits an
approximately quadratic dependence on $B$ in the weak-field
regime. Its sign and magnitude, however, depend sensitively on
the monopole charge and the nature of impurity scattering.

For $N=1$, increasing the intervalley scattering parameter
$\alpha$ progressively suppresses the magnitude of the PHC,
while preserving its positive sign and approximately quadratic
field dependence throughout the investigated parameter range.
This suppression is consistent with the weakening of the
positive anomaly-assisted contribution discussed for the LMC:
stronger intervalley scattering more efficiently relaxes the
nonequilibrium chiral imbalance generated by the
$\mathbf{E}\cdot\mathbf{B}$ pumping term. The associated current
flows along $\mathbf{B}$ and therefore contributes to both the
longitudinal and transverse in-plane responses when the magnetic
field is oblique to the electric field. Its transverse component
has the characteristic dependence
$B_x(\mathbf{E}\cdot\mathbf{B})\propto B^2\sin\gamma\cos\gamma$.
Consequently, increasing $\alpha$ reduces this positive
contribution without changing its even symmetry under
magnetic-field reversal.

\textcolor{black}{
For $N=1$, the $\sigma_x$ and $\sigma_y$ impurity channels,
shown in Fig.~\ref{Fig:N123_PHC_vs_B_alp_vary_OMM1.png}(d)
and (g), respectively, generally yield a larger PHC than
the nonmagnetic $\sigma_0$ and magnetic $\sigma_z$ channels
at the same intervalley scattering strength. This impurity
dependence originates from the different pseudospin selection
rules entering the collision integral. As expressed by the
squared overlap in Eq.~\ref{eq:app-overlap}, conjugation by
$\sigma_i$, with $i=x,y,z$, preserves the $i$ component of
the pseudospin vector and reverses its other two components,
whereas $\sigma_0$ preserves all three. Consequently, the
different impurity potentials assign different weights to
scattering between the same initial and final states.
For multi-Weyl fermions, these weights depend on the
momentum-dependent pseudospin texture and cannot, in general,
be represented by a single impurity-dependent sign multiplying
$\mathbf{k}\cdot\mathbf{k}'$.
}

\textcolor{black}{
This distinction can be illustrated by intervalley scattering
between states on opposite poles of the zero-field Fermi
surfaces, with the initial momentum along $+\hat{\mathbf{z}}$
and the final momentum along $-\hat{\mathbf{z}}$, measured
relative to their respective nodes. For conduction-band states
with $\chi'=-\chi$, the pseudospins of these two states are
parallel. The normalized squared matrix element is therefore
unity for $\sigma_0$ and $\sigma_z$, but vanishes for
$\sigma_x$ and $\sigma_y$. The latter channels thus suppress
this particular intervalley backscattering process, illustrating
how the impurity pseudospin structure can favor a larger
transport response. Here, the distinction between intra- and
intervalley scattering is specified by $\chi$ and $\chi'$;
the Pauli matrix determines the pseudospin matrix element
and does not independently change the node chirality.
At the oblique field orientation considered here, the PHC
depends on scattering over the entire OMM-modified Fermi
surface. The pole-to-pole selection rule therefore provides
a microscopic illustration, while the quantitative response
requires the full momentum-dependent collision integral.
}

A striking feature of
Fig.~\ref{Fig:N123_PHC_vs_B_alp_vary_OMM1.png} is the
substantially larger PHC magnitude in the $N=2$ and $N=3$
systems compared with $N=1$, within the investigated parameter
regime. As discussed for the LMC, this enhancement reflects
the combined influence of monopole charge, anisotropic band
dispersion, and impurity scattering. Increasing $N$ modifies
the magnitude and momentum-space distribution of the Berry
curvature and OMM, together with the Fermi surface, carrier
velocities, and scattering matrix elements entering the
transport integrals. These changes can enhance the transverse
response arising from the Berry-curvature-modified dynamics
and OMM-dependent corrections. This behavior is consistent
with theoretical studies demonstrating a strong dependence
of multi-Weyl magnetotransport on monopole charge
\cite{huang2017topological,dantas2018magnetotransport,nag2021magneto}.
The magnitude of this enhancement also depends on the
dispersion coefficients, Fermi energy, and scattering channel;
it should therefore not be interpreted as a universal
monotonic scaling with $N$ determined by topology alone.

\textcolor{black}{
Beyond the enhancement in magnitude, the double- and
triple-Weyl systems exhibit a qualitative distinction from
the single-Weyl case: the PHC undergoes a sign reversal at
sufficiently strong intervalley scattering for the $\sigma_0$
and $\sigma_z$ impurity channels.
}
As shown in
Fig.~\ref{Fig:N123_PHC_vs_B_alp_vary_OMM1.png}(b), (c), (k),
and (l), increasing $\alpha$ changes the response from positive
to negative, while preserving its approximately quadratic
magnetic-field dependence and even symmetry under
$B\rightarrow -B$. The transition therefore corresponds to
a change in the sign of the quadratic coefficient, analogous
to the LMC sign reversal discussed above. However, whereas
the LMC exhibits this transition for all three monopole
charges in these impurity channels, the PHC remains positive
for $N=1$ throughout the investigated parameter range.
Thus, the longitudinal and planar Hall responses need not
undergo sign reversal under the same conditions.

Consistent with the LMC discussion, this transition can be
understood in terms of the competition between the positive
anomaly-assisted planar Hall response and the corrections
arising from the combined effects of Berry curvature, OMM,
and impurity scattering. Increasing the intervalley scattering
strength reduces the chiral imbalance and suppresses the
associated positive contribution. Meanwhile, the OMM modifies
the quasiparticle dispersion, Fermi surface, and carrier
velocities, thereby changing the scattering phase space and
the nonequilibrium distribution. Together with the
Berry-curvature-modified dynamics, these changes determine
the magnitude and sign of the quadratic planar Hall response.
The negative PHC should therefore not be attributed to an
independently negative OMM-only contribution. For the
$\sigma_0$ and $\sigma_z$ channels in the $N=2$ and $N=3$
systems, the observed sign reversal is consistent with these
combined corrections overcoming the positive anomaly-assisted
contribution at sufficiently strong intervalley scattering.

\textcolor{black}{
In contrast, the PHC remains positive for the $\sigma_x$ and
$\sigma_y$ impurity channels in the double- and triple-Weyl
systems throughout the displayed range, as shown in
Fig.~\ref{Fig:N123_PHC_vs_B_alp_vary_OMM1.png}(e), (f), (h),
and (i). These channels also support a larger positive response
over much of the investigated range. Their different overlap
factors redistribute the angular weights of intra- and
intervalley scattering and consequently modify the
nonequilibrium distribution. The absence of a sign reversal
demonstrates that the band geometry associated with higher
monopole charge does not, by itself, guarantee negative PHC.
As for the LMC, the impurity structure of the collision
integral is essential in determining whether the total
quadratic response changes sign.
}

The emergence of negative PHC in the double- and triple-Weyl
systems is also relevant to the broader comparison between
higher-charge and higher-pseudospin Weyl fermions. A similar
transition from positive to negative PHC has been predicted
for pseudospin-$1$ Weyl semimetals when intervalley scattering
is increased in the presence of OMM
\cite{ahmad2026chiral}. In that setting, the dispersive bands
carry Chern numbers of magnitude $|C|=2$, and their band
geometry arises from the enlarged pseudospin representation.
The present multi-Weyl systems instead remain two-band,
pseudospin-$1/2$ systems, with higher monopole charge encoded
in the momentum-space winding and nonlinear transverse
dispersion. Our results show that negative PHC can occur
in both settings. This correspondence suggests that the
phenomenon is not restricted to a particular pseudospin
representation, although its occurrence and threshold depend
on the specific band structure and scattering processes.

Taken together, these results identify the PHC as a sensitive
probe of the interplay between Berry curvature, OMM,
anisotropic dispersion, and impurity scattering. While the
single-Weyl system retains a positive quadratic PHC over the
investigated range, the double- and triple-Weyl systems can
develop a negative quadratic response in selected impurity
channels. In conjunction with studies of higher-pseudospin
transport \cite{ahmad2025longitudinal,ahmad2026chiral}, these
findings motivate a comparison of how band geometry and
scattering jointly control magnetotransport across different
classes of Weyl systems. Here, the sign reversal concerns
the coefficient of the response at order $B^2$ within
linear response to the electric field.

\begin{figure*}
\centering
\includegraphics[width=2\columnwidth]{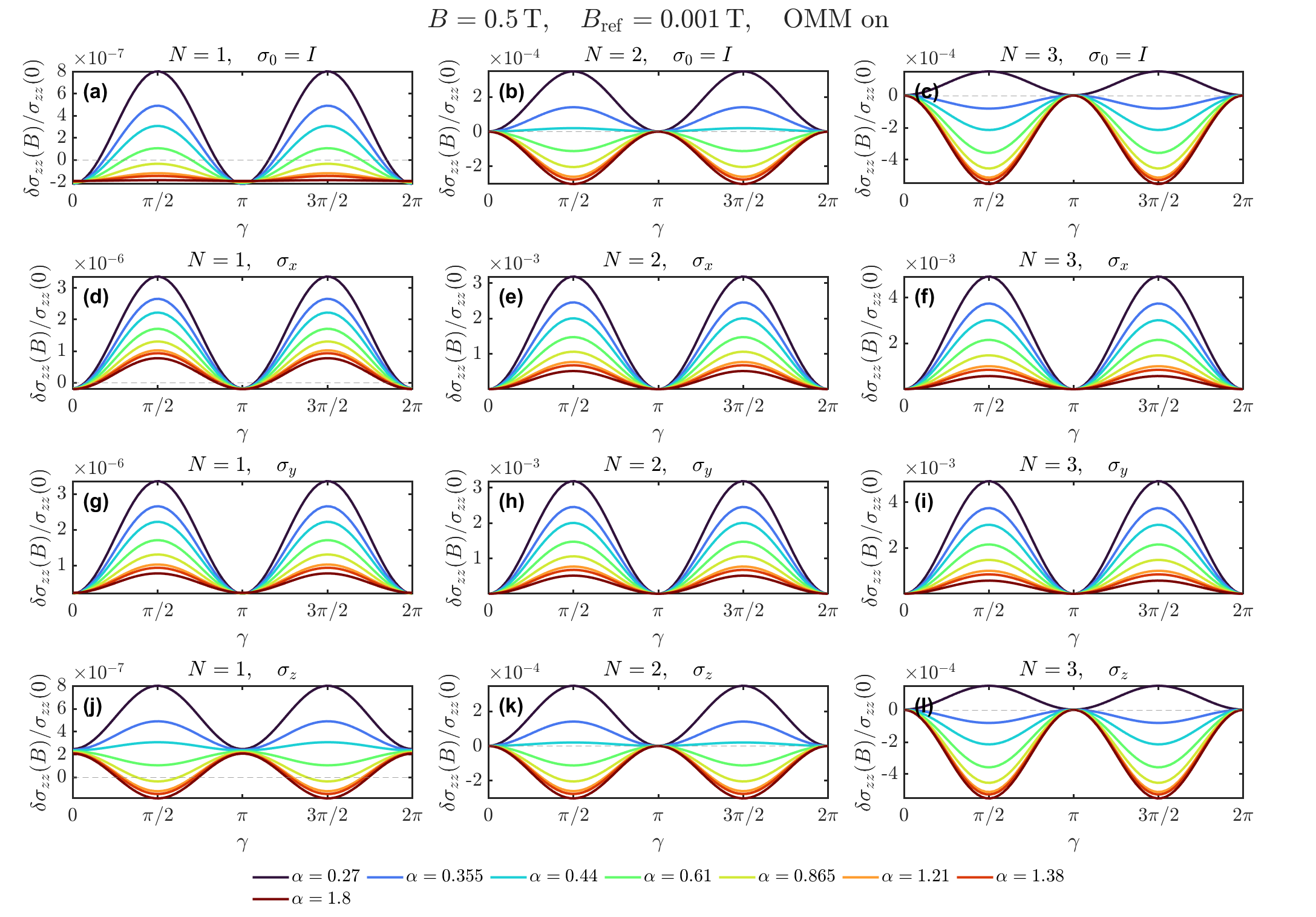}
\caption{LMC as a function of the magnetic-field orientation angle $\gamma$ for different values of the intervalley scattering parameter $\alpha$ in multi-Weyl semimetals with topological charge (a) $N=1$, (b) $N=2$, and (c) $N=3$. The orbital OMM contribution is included in all panels. For $N=1$, the LMC exhibits the characteristic $\sin^2\gamma$ angular dependence expected for chiral-anomaly-assisted transport. For a fixed $\gamma$, increasing $\alpha$ continuously suppresses the positive LMC and eventually drives a sign reversal, resulting in a negative response. For $N=2$ and $N=3$, while the angular profile remains sinusoidal throughout, unlike the conventional Weyl-semimetal case, no finite OMM-induced vertical offset appears in the multi-Weyl systems, and the LMC always vanishes at $\gamma=0,\pi$, and $2\pi$, it shows a trasition from $\sin^2\gamma$ to $-\sin^2\gamma$ above a critical value of $\alpha$. The overall magnitude of the angular magnetoconductivity increases significantly with increasing topological charge $N$.} 
\label{Fig:N123_LMC_vs_gm_alp_vary_OMM1.png}
\end{figure*}
\begin{figure*}
\centering
\includegraphics[width=2\columnwidth]{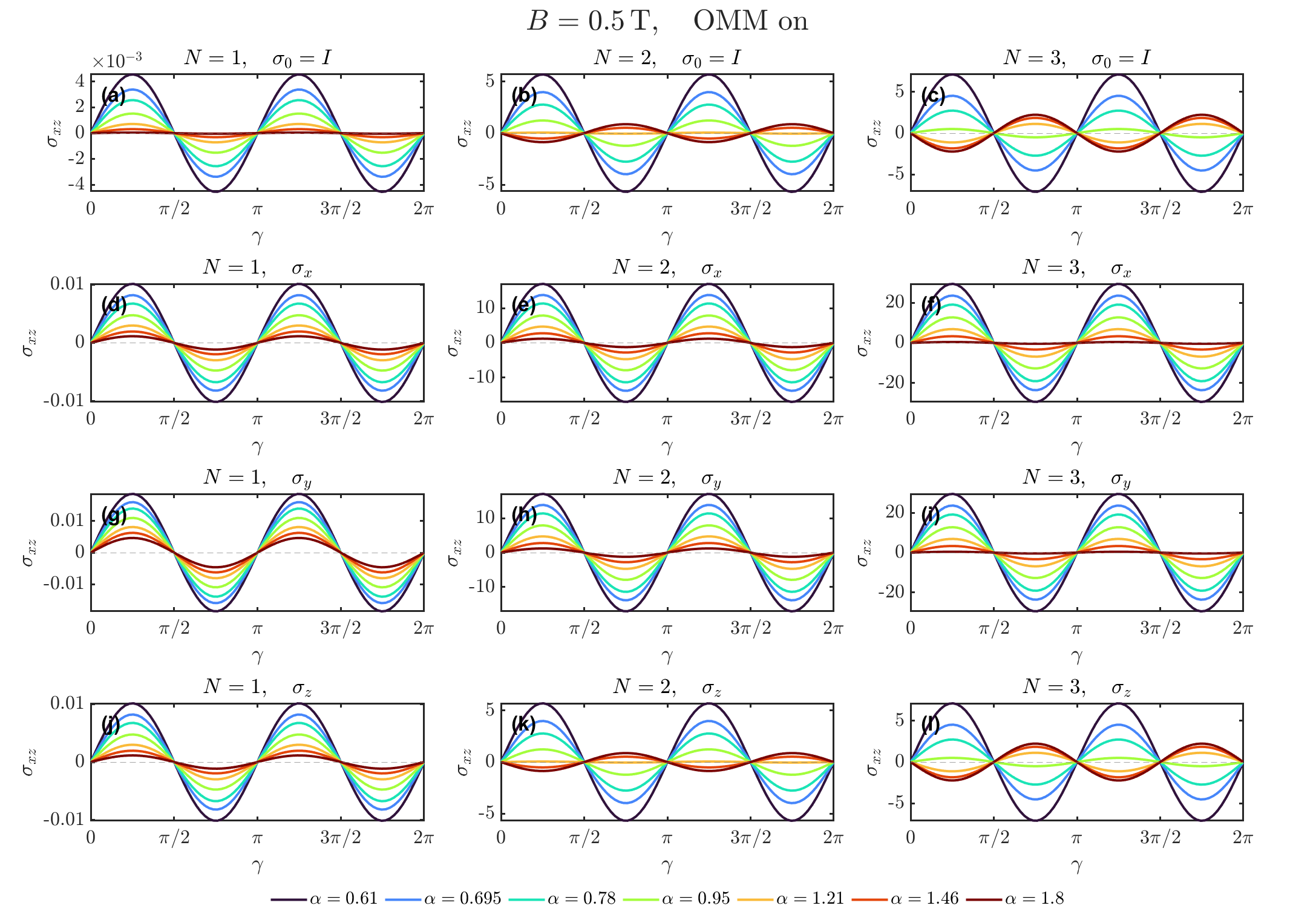}
\caption{PHC, $\sigma_{zx}(B)$, as a function of the
magnetic-field orientation angle $\gamma$ for
different intervalley scattering strengths $\alpha$
at $B=0.5$~T, with OMM included.
The left, middle, and right columns correspond
to monopole charges $N=1$, $2$, and $3$,
respectively. Within each column, the four
impurity potentials $\sigma_0=I$, $\sigma_x$,
$\sigma_y$, and $\sigma_z$ are shown in
successive rows. The PHC exhibits an approximately
$\sin2\gamma$ angular dependence. Increasing
$\alpha$ progressively suppresses its amplitude
and induces a sign reversal for $N=2$ and $N=3$
in the $\sigma_0$ and $\sigma_z$ channels.
The $\sigma_x$ and $\sigma_y$ channels retain
a positive $\sin2\gamma$ coefficient throughout
the investigated range.}
\label{Fig:N123_PHC_vs_gm_alp_vary_OMM1.png}
\end{figure*}
\subsection{Angular dependence of LMC and PHC}
\label{Sec.:Angular dependence of LMC and PHC}
To further elucidate the interplay between Berry curvature, orbital magnetic moment (OMM), and intervalley scattering, we now examine the angular dependence of the LMC and PHC. The magnetic field is rotated in the $xz$ plane according to $\mathbf{B}=B(\cos\gamma,0,\sin\gamma)$, while the electric field is kept fixed along the $z$ direction, $\mathbf{E}=E\hat{\mathbf{z}}$. In this configuration, the relative orientation of the electric and magnetic fields is controlled by $\gamma$, allowing us to investigate how the combined effects of Berry curvature, OMM, and momentum-dependent impurity scattering influence the angular dependence and sign of the magnetotransport response in multi-Weyl semimetals.

Figure~\ref{Fig:N123_LMC_vs_gm_alp_vary_OMM1.png} shows the angular dependence of the normalized LMC at a fixed magnetic field $B=0.5$~T for monopole charges $N=1$, $2$, and $3$, with the OMM included in all calculations. In the weak intervalley scattering regime, the LMC remains predominantly positive and approximately follows the characteristic $\sin^{2}\gamma$ dependence associated with chiral-anomaly-assisted transport. The response approaches a small finite value at $\gamma=0$, $\pi$, and $2\pi$, where the magnetic field is perpendicular to the electric field, and reaches its maximum magnitude near $\gamma=\pi/2$ and $3\pi/2$, corresponding to parallel and antiparallel field configurations, respectively. This angular behavior is consistent with the anomaly-induced charge pumping proportional to $\mathbf{E}\cdot\mathbf{B}$, which generates a positive longitudinal magnetoconductivity varying as $B^2\sin^2\gamma$ in the conventional weak-field description~\cite{son2012chiral,dantas2018magnetotransport}. The persistence of a similar angular profile in the present calculations indicates that the combined effects of Berry curvature, OMM, and momentum-dependent scattering approximately preserve the characteristic angular symmetry of the positive LMC in this regime.

The inclusion of the OMM introduces a remarkable qualitative change in the angular dependence of the LMC. At a fixed angle $\gamma$, increasing the intervalley scattering strength $\alpha$ progressively suppresses the positive magnetoconductivity and can eventually drive a transition to negative LMC. Importantly, this sign reversal occurs without significantly altering the characteristic angular symmetry of the response. As shown in Fig.~\ref{Fig:N123_LMC_vs_gm_alp_vary_OMM1.png}, the angular profiles retain an approximately $\sin^{2}\gamma$ dependence with a small offset, while their amplitudes decrease and eventually become negative for sufficiently strong intervalley scattering in the relevant impurity channels. This behavior is particularly pronounced for $N=2$ and $N=3$ in the $\sigma_0$ and $\sigma_z$ channels, whereas the $\sigma_x$ and $\sigma_y$ channels retain a positive response throughout the investigated range. Thus, intervalley scattering, together with the OMM-dependent geometric corrections, primarily modifies the magnitude and sign of the LMC while approximately preserving its characteristic angular symmetry within the considered parameter regime.

A notable feature of the angular LMC is the presence of a small angle-independent contribution in addition to the dominant $\sin^{2}\gamma$ dependence. A closer examination of the numerical results, including the zero-field subtraction and angular-grid convergence, confirms that the response does not strictly vanish for perpendicular electric and magnetic fields. This contribution is relatively weak for $N=2$ and $N=3$, where the angular dependence is dominated by the $\sin^{2}\gamma$ term, whereas it becomes more noticeable for $N=1$ and depends on the impurity channel. The OMM-dependent modifications of the quasiparticle dispersion and velocity, together with Berry-curvature-modified dynamics and momentum-dependent scattering, influence both the angular variation and the overall magnitude of the LMC. Consequently, the transition from positive to negative magnetoconductivity is primarily associated with a change in the coefficient of the $\sin^{2}\gamma$-like response, while the underlying angular symmetry remains approximately preserved within the investigated parameter regime.

The enhancement of the angular LMC with increasing topological charge is another striking feature of Fig.~\ref{Fig:N123_LMC_vs_gm_alp_vary_OMM1.png}. The magnitude of both the positive and negative responses grows substantially as $N$ increases from $1$ to $3$ within the investigated parameter regime. This behavior is associated with the modified Berry curvature and orbital magnetic moment distributions of higher-order Weyl nodes, while the anomaly coefficient itself is proportional to the monopole charge. The combined geometric effects and anisotropic scattering processes therefore enhance the sensitivity of the conductivity to intervalley scattering, making the sign reversal increasingly pronounced in higher-order multi-Weyl semimetals~\cite{huang2017topological,dantas2018magnetotransport}.

The angular dependence of the PHC is presented in Fig.~\ref{Fig:N123_PHC_vs_gm_alp_vary_OMM1.png} for monopole charges $N=1$, $2$, and $3$, with the OMM included in all calculations. In the weak-to-intermediate intervalley scattering regime, the PHC exhibits approximately the characteristic $\sin2\gamma$ dependence expected for planar Hall transport in systems where the electric and magnetic fields lie in the same plane. The conductivity vanishes at $\gamma=0$, $\pi/2$, $\pi$, $3\pi/2$, and $2\pi$, and reaches extrema near $\gamma=\pi/4$, $3\pi/4$, $5\pi/4$, and $7\pi/4$, reflecting the transverse response associated with the in-plane components of the magnetic field. Increasing $\alpha$ progressively suppresses the magnitude of the PHC, while increasing the topological charge generally enhances its amplitude within the investigated parameter regime. The persistence of the approximately $\sin2\gamma$ angular dependence is consistent with the conventional Berry-curvature-mediated planar Hall response, even in the presence of OMM and intervalley scattering~\cite{sharma2020sign,ahmad2021longitudinal}.

A qualitatively different behavior emerges in the strong intervalley scattering regime. For $N=1$, the coefficient of the $\sin2\gamma$ response remains positive throughout the investigated range, although its magnitude becomes substantially reduced. In contrast, both the double- and triple-Weyl systems undergo a sign reversal for the $\sigma_0$ and $\sigma_z$ impurity channels, as shown in Figs.~\ref{Fig:N123_PHC_vs_gm_alp_vary_OMM1.png}(b), (c), (k), and (l). The PHC evolves from an approximately $+\sin2\gamma$ profile to a $-\sin2\gamma$ profile as $\alpha$ increases, with the transition occurring at a smaller intervalley scattering strength for $N=3$ than for $N=2$. Meanwhile, the $\sigma_x$ and $\sigma_y$ channels retain a positive $\sin2\gamma$ coefficient throughout the investigated range. Importantly, the sign reversal occurs without significantly altering the characteristic angular symmetry of the PHC. This behavior suggests that the combined effects of Berry curvature and OMM, together with momentum-dependent scattering, become sufficiently important to overcome the positive anomaly-assisted contribution in the higher-charge Weyl systems. Thus, the sign reversal depends sensitively on both the monopole charge and the nature of impurity scattering.

The sign reversal of the PHC in the $N=2$ and $N=3$ systems closely parallels the behavior recently predicted for pseudospin-1 Weyl semimetals, where the PHC changes from positive to negative under sufficiently strong intervalley scattering in the presence of OMM~\cite{ahmad2026chiral}. In pseudospin-1 semimetals, the enhanced geometric response is associated with higher-pseudospin quasiparticles whose dispersive bands carry Chern numbers $|C|=2$. The present results demonstrate that a similar transition occurs in multi-Weyl semimetals with topological charges $N=2$ and $N=3$, even though the underlying quasiparticles remain pseudospin-$1/2$. In both systems, the interplay of Berry curvature, OMM, and impurity scattering can substantially modify the planar Hall response. However, unlike the pseudospin-1 case, where the band geometry is associated with the enlarged pseudospin representation, the present behavior is influenced by the higher monopole charge and nonlinear transverse dispersion of multi-Weyl nodes. This correspondence suggests that negative PHC under strong intervalley scattering can arise in distinct classes of topological semimetals rather than being restricted to a particular pseudospin representation.

Taken together, Figs.~\ref{Fig:N123_LMC_vs_gm_alp_vary_OMM1.png} and \ref{Fig:N123_PHC_vs_gm_alp_vary_OMM1.png} demonstrate that the angular dependences of LMC and PHC retain their characteristic $\sin^{2}\gamma$-like and $\sin2\gamma$ forms, respectively, even when the magnetotransport response undergoes a sign reversal. While the LMC contains a small additional angle-independent contribution, the PHC remains approximately proportional to $\sin2\gamma$ within the investigated parameter regime. Increasing intervalley scattering suppresses the positive anomaly-assisted response, while the combined OMM-related geometric and scattering corrections can drive the LMC and PHC negative in the relevant impurity channels. The pronounced sign reversal for higher monopole charges highlights the importance of anisotropic band geometry, orbital magnetic moment, and intervalley scattering in determining magnetotransport in multi-Weyl semimetals.

\section{Conclusions}
\label{Sec:Conclusions}

In this work, we have investigated the LMC and PHC in multi-Weyl semimetals with monopole charges $N=1$, $2$, and $3$, within a semiclassical Boltzmann transport framework beyond the constant relaxation-time approximation. By incorporating Berry curvature and orbital magnetic moment (OMM) effects into the quasiparticle dynamics and treating intra- and intervalley scattering through a momentum-dependent collision integral, we have systematically examined the influence of band geometry, monopole charge, and impurity scattering on the magnetotransport response. Both nonmagnetic and magnetic impurity channels are considered, allowing us to identify the role of their distinct scattering structures in determining the sign and magnitude of the conductivity.

Our results demonstrate that the OMM plays a crucial role in the emergence of negative LMC. In its absence, the LMC remains positive and approximately quadratic in the magnetic field throughout the investigated intervalley scattering range, consistent with conventional chiral-anomaly-assisted transport. Incorporating OMM, however, produces a sign reversal at sufficiently strong intervalley scattering for the $\sigma_0$ and $\sigma_z$ impurity channels. Remarkably, the critical scattering strength decreases systematically with increasing monopole charge, from $\alpha_c\simeq0.7806$ for $N=1$ to $0.4613$ and $0.323$ for $N=2$ and $N=3$, respectively. The negative quadratic LMC also becomes more pronounced for higher-charge nodes within the investigated parameter regime. In contrast, the $\sigma_x$ and $\sigma_y$ channels retain positive LMC, demonstrating that the sign reversal depends sensitively on the impurity scattering mechanism. These findings extend the weak-field negative-LMC behavior established for conventional Weyl semimetals to higher-charge multi-Weyl systems and highlight the combined influence of Berry curvature, OMM, and momentum-dependent scattering~\cite{knoll2020negative}.

The planar Hall response exhibits a distinct dependence on
monopole charge and impurity scattering. With OMM included,
the PHC remains positive for $N=1$ throughout the investigated
parameter range, whereas sufficiently strong intervalley
scattering drives a sign reversal for $N=2$ and $N=3$ in
the $\sigma_0$ and $\sigma_z$ impurity channels. In contrast,
the $\sigma_x$ and $\sigma_y$ channels retain a positive
PHC over the same range. The response remains approximately
quadratic in magnetic field and even under field reversal,
so the transition corresponds to a change in the sign of
its quadratic coefficient. Unlike the LMC, which reverses
sign for all three monopole charges in the $\sigma_0$ and
$\sigma_z$ channels, the PHC therefore exhibits different
sign-reversal conditions. These findings emphasize that
negative PHC arises from the combined effects of band
geometry and impurity scattering, rather than from an
independently negative OMM contribution.

Overall, our findings establish that the sign of magnetoconductivity in multi-Weyl semimetals is not determined by the chiral anomaly or monopole charge alone, but results from the combined effects of band geometry, OMM, and the momentum-dependent intra- and intervalley scattering processes. In particular, higher monopole charges can substantially enhance the tendency toward negative LMC and PHC for specific impurity channels, despite the positive anomaly-assisted contribution. The predicted dependence of these sign reversals on monopole charge and impurity scattering provides a useful theoretical framework for interpreting weak-field magnetotransport measurements and identifying the role of geometric corrections in multi-Weyl semimetals.
\section{Acknowledgments}
A.A. would like to acknowledge financial support from
Indian Institute of Technology Bombay through Institute Post
Doctoral Fellowship. S.M. acknowledges financial support from the Anusandhan National Research Foundation (ANRF), India through the National Postdoctoral Fellowship
(PDF/2025/003070). G.V.K. acknowledges financial support from ANRF-SERB
Core Research Grant CRG/2023/005628. M.H.R. don't get any financial support. T.T. acknowledges financial support by the Japan Society for the Promotion of Science, KAKENHI (Grants No. 25H01248) from the Ministry of Education, Culture, Sports, Science, and Technology, Japan. The authors are grateful to Gargee Sharma for fruitful discussions.
\appendix
\begin{widetext}
\section{Formation of system of linear equations from Maxwell-Boltzmann transport theory}
For a single impurity channel,$ U_{\mathbf k\mathbf k'}^{\chi\chi'}=U^{\chi\chi'}\sigma_i,
 i=0,x,y,z,\sigma_0=I_{2\times2}$. One can define a projector associated with the conduction band at a node as, $P_\chi=\frac{I+\chi\,\mathbf u^\chi\cdot\boldsymbol\sigma}{2}$. Here, $\mathbf u^\chi$ is a vector with components defined in the main text. Consequently, the squared overlap of the conduction wave-function amplitude is
\begin{align}
\mathcal O_i^{\chi\chi'}&=
\left|\langle\psi_{\chi'}(\mathbf k')|\sigma_i|\psi_\chi(\mathbf k)\rangle\right|^2
=\operatorname{Tr}\!\left(P_{\chi'}\sigma_iP_\chi\sigma_i\right)\\
&=\frac12\left[1+\chi\chi'\sum_{a=x,y,z}C_{i,a}u_a^\chi u_a'\right],
\qquad
\mathcal T_i^{\chi\chi'}=|U^{\chi\chi'}|^2\mathcal O_i^{\chi\chi'}.
\label{eq:app-overlap}
\end{align}
For $i=x,y,z$, conjugation by $\sigma_i$ preserves the $i$ component of a Pauli vector and reverses the other two. The signs are listed in Table~\ref{tab:app-signs}; the impurity label $i$ is distinct from the node sign $\chi$ and the winding number $N$. Written explicitly in the original angular variables, the overlap is
\begin{equation}
\mathcal T_i^{\chi\chi'}=\frac{|U^{\chi\chi'}|^2}{2}
\left[1+\chi\chi'\frac{C_{i,z}\beta^\chi\beta'\cos\theta\cos\theta'
+\sin^N\theta\sin^N\theta'[C_{i,x}\cos(N\phi)\cos(N\phi')+C_{i,y}\sin(N\phi)\sin(N\phi')]}{R^\chi R'}\right].
\label{eq:app-overlap-explicit}
\end{equation}
To solve the radial energy integration, we have used 
\begin{equation}
\left|\frac{\partial\widetilde\epsilon^\chi}{\partial k}\right|
=\hbar|\mathbf v^\chi\cdot\widehat{\mathbf k}|,
\qquad
\int_0^\infty k^2dk\,\delta(\widetilde\epsilon^\chi-\epsilon_F)X(k)
=\left.\frac{k^3X(k)}{\hbar|\mathbf v^\chi\cdot\mathbf k|}\right|_{k=k_F^\chi}.
\end{equation}
Following the main-text notation, define
\begin{equation}
\mathcal F^\chi(\theta,\phi)=
\left.\frac{k^3\sin\theta}{|\mathbf v^\chi\cdot\mathbf k|}
(\mathcal D^\chi)^{-1}\right|_{k=k_F^\chi}.
\label{eq:app-Fmeasure}
\end{equation}
Incorporating Fermi's golden rule, the chiral scattering rates are
\begin{equation}
\frac{1}{\tau_i^\chi}=\sum_{\chi'}\frac{2\pi n_{\rm imp}}{\hbar}
\int\frac{d^3k'}{(2\pi)^3}(\mathcal D^{\chi'})^{-1}
|U^{\chi\chi'}|^2\mathcal O_i^{\chi\chi'}
\delta(\widetilde\epsilon^{\chi'}-\epsilon_F).
\end{equation}
We define the following substitutions, $\kappa^{\chi\chi'}=\frac{n_{\rm imp}|U^{\chi\chi'}|^2}{8\pi^2\hbar^2},
K_i^{\chi\chi'}=1+\chi\chi'\sum_a C_{i,a}u_a^\chi u_a'$, the scattering rate is written as
\begin{align}
\frac{1}{\tau_i^\chi(\theta,\phi)}&=
\sum_{\chi'}\kappa^{\chi\chi'}\int d\theta'd\phi'\,
\mathcal F^{\chi'}(\theta',\phi')K_i^{\chi\chi'}.
\label{eq:app-rate}
\end{align}
Equivalently, when the kernel is $\mathcal T_i=|U|^2K_i/2$, its prefactor is $n_{\rm imp}/(4\pi^2\hbar^2)$ with these continuum conventions. Defining a handful of stuff:
\begin{equation}
\overline F^\chi=\int\mathcal F^\chi d\theta d\phi,
\qquad
(\overline G^\chi,\overline I^\chi,\overline J^\chi)
=\int\mathcal F^\chi(u_z^\chi,u_x^\chi,u_y^\chi)d\theta d\phi.
\end{equation}
Equation~\eqref{eq:app-rate} then becomes
\begin{equation}
(\tau_i^\chi)^{-1}=\sum_{\chi'}\kappa^{\chi\chi'}
\left[\overline F^{\chi'}+\chi\chi'
\left(C_{i,z}u_z^\chi\overline G^{\chi'}+
C_{i,x}u_x^\chi\overline I^{\chi'}+C_{i,y}u_y^\chi\overline J^{\chi'}\right)\right].
\label{eq:app-rate-mom}
\end{equation}
The external factors $u_a^\chi$ always belong to the state whose lifetime is being calculated, in both intranode and internode terms. Please note that the subscript $i$ incorporates the magnetic as well as non-magnetic scattering centers. The integrated moments belong to $\chi'$.
For $\mathbf E=E\hat{\mathbf z}$ and $\mathbf B=B(\cos\gamma,0,\sin\gamma)$, write;$ g^\chi_{\mathbf k}=-eE\frac{\partial f_0}{\partial\widetilde\epsilon^\chi}\Lambda^\chi,
h^\chi=\mathcal D^\chi\left[v_z^\chi+\frac{eB\sin\gamma}{\hbar}(\boldsymbol\Omega^\chi\cdot\mathbf v^\chi\right)]$.

The Boltzmann transport-collision equation reads
\begin{equation}
h^\chi+\frac{\Lambda^\chi}{\tau_i^\chi}
=\sum_{\chi'}\kappa^{\chi\chi'}\int d\theta'd\phi'\,
\mathcal F^{\chi'}K_i^{\chi\chi'}\Lambda^{\chi'}.
\label{eq:app-BTE}
\end{equation}
In principle, the kernel is separable in the four external functions $1,u_z^\chi,u_x^\chi,u_y^\chi$. Therefore, one can choose the following ansatz
\begin{equation}
\Lambda_i^\chi=\tau_i^\chi
\left[d^\chi-h^\chi+a^\chi u_z^\chi+b^\chi u_x^\chi+c^\chi u_y^\chi\right].
\label{eq:app-ansatz}
\end{equation}
Suppress the fixed impurity label on $\tau_i^\chi$ and define
\begin{equation}
\langle X\rangle_\chi=\int d\theta d\phi\,
\mathcal F^\chi(\theta,\phi)\tau_i^\chi(\theta,\phi)X(\theta,\phi).
\label{eq:app-average}
\end{equation}
All unbarred quantities below use this same weight:
\begin{align}
F^\chi&=\langle1\rangle_\chi,&G^\chi&=\langle u_z^\chi\rangle_\chi,&I^\chi&=\langle u_x^\chi\rangle_\chi,&J^\chi&=\langle u_y^\chi\rangle_\chi,\\
O^\chi&=\langle(u_z^\chi)^2\rangle_\chi,&P^\chi&=\langle u_z^\chi u_x^\chi\rangle_\chi,&Q^\chi&=\langle u_z^\chi u_y^\chi\rangle_\chi,\\
S^\chi&=\langle(u_x^\chi)^2\rangle_\chi,&U^\chi&=\langle u_x^\chi u_y^\chi\rangle_\chi,&V^\chi&=\langle(u_y^\chi)^2\rangle_\chi,\\
H^\chi&=\langle h^\chi\rangle_\chi,&N^\chi&=\langle h^\chi u_z^\chi\rangle_\chi,&L^\chi&=\langle h^\chi u_x^\chi\rangle_\chi,&M^\chi&=\langle h^\chi u_y^\chi\rangle_\chi.
\label{eq:app-moments}
\end{align}
Please note that $N^\chi$ is not the winding number $N$, and is defined locally. For example,
\begin{align}
G^\chi&=\int d\theta d\phi\,\mathcal F^\chi\tau_i^\chi
\frac{\beta^\chi\cos\theta}{R^\chi},\\
O^\chi&=\int d\theta d\phi\,\mathcal F^\chi\tau_i^\chi
\frac{(\beta^\chi)^2\cos^2\theta}{(R^\chi)^2},
&P^\chi&=\int d\theta d\phi\,\mathcal F^\chi\tau_i^\chi
\frac{\beta^\chi\cos\theta\sin^N\theta\cos(N\phi)}{(R^\chi)^2}.
\end{align}

Substituting Eq.~\eqref{eq:app-ansatz} into Eq.~\eqref{eq:app-BTE} and matching coefficients of the same angular function from both sides,
\begin{align}
d^\chi&=\sum_{\chi'}\kappa^{\chi\chi'}
\left(F^{\chi'}d^{\chi'}+G^{\chi'}a^{\chi'}+I^{\chi'}b^{\chi'}+J^{\chi'}c^{\chi'}-H^{\chi'}\right),\\
a^\chi&=\sum_{\chi'}\kappa^{\chi\chi'}\chi\chi'C_{i,z}
\left(G^{\chi'}d^{\chi'}+O^{\chi'}a^{\chi'}+P^{\chi'}b^{\chi'}+Q^{\chi'}c^{\chi'}-N^{\chi'}\right),\\
b^\chi&=\sum_{\chi'}\kappa^{\chi\chi'}\chi\chi'C_{i,x}
\left(I^{\chi'}d^{\chi'}+P^{\chi'}a^{\chi'}+S^{\chi'}b^{\chi'}+U^{\chi'}c^{\chi'}-L^{\chi'}\right),\\
c^\chi&=\sum_{\chi'}\kappa^{\chi\chi'}\chi\chi'C_{i,y}
\left(J^{\chi'}d^{\chi'}+Q^{\chi'}a^{\chi'}+U^{\chi'}b^{\chi'}+V^{\chi'}c^{\chi'}-M^{\chi'}\right).
\label{eq:app-coefficients}
\end{align}
We list the unknowns in the following order
\begin{equation}
\mathbb Z=(d^+,a^+,b^+,c^+,d^-,a^-,b^-,c^-)^T,
\qquad (\mathbb A-\mathbb I_8)\mathbb Z=\mathbb Y.
\label{eq:app-system}
\end{equation}
Introduce
\begin{align}
\mathsf M_\chi&=\begin{pmatrix}
F^\chi&G^\chi&I^\chi&J^\chi\\
G^\chi&O^\chi&P^\chi&Q^\chi\\
I^\chi&P^\chi&S^\chi&U^\chi\\
J^\chi&Q^\chi&U^\chi&V^\chi
\end{pmatrix},
&\mathbf H_\chi&=\begin{pmatrix}H^\chi\\N^\chi\\L^\chi\\M^\chi\end{pmatrix},\\
\mathsf S_+&=\operatorname{diag}(1,C_{i,z},C_{i,x},C_{i,y}),
&\mathsf S_-&=\operatorname{diag}(1,-C_{i,z},-C_{i,x},-C_{i,y}).
\end{align}
Then
\begin{equation}
\mathbb A=\begin{pmatrix}
\kappa^{++}\mathsf S_+\mathsf M_+&\kappa^{+-}\mathsf S_-\mathsf M_-\\
\kappa^{-+}\mathsf S_-\mathsf M_+&\kappa^{--}\mathsf S_+\mathsf M_-
\end{pmatrix},\qquad
\mathbb Y=\begin{pmatrix}
\kappa^{++}\mathsf S_+\mathbf H_++\kappa^{+-}\mathsf S_-\mathbf H_-\\
\kappa^{-+}\mathsf S_-\mathbf H_++\kappa^{--}\mathsf S_+\mathbf H_-
\end{pmatrix}.
\label{eq:app-blocks}
\end{equation}
For complete specification, each block has the form
\begin{equation}
\mathbb A_{\chi\chi'}=\kappa^{\chi\chi'}
\begin{pmatrix}
F^{\chi'}&G^{\chi'}&I^{\chi'}&J^{\chi'}\\
\chi\chi'C_{i,z}G^{\chi'}&\chi\chi'C_{i,z}O^{\chi'}&\chi\chi'C_{i,z}P^{\chi'}&\chi\chi'C_{i,z}Q^{\chi'}\\
\chi\chi'C_{i,x}I^{\chi'}&\chi\chi'C_{i,x}P^{\chi'}&\chi\chi'C_{i,x}S^{\chi'}&\chi\chi'C_{i,x}U^{\chi'}\\
\chi\chi'C_{i,y}J^{\chi'}&\chi\chi'C_{i,y}Q^{\chi'}&\chi\chi'C_{i,y}U^{\chi'}&\chi\chi'C_{i,y}V^{\chi'}
\end{pmatrix}.
\end{equation}
The fully expanded driving vector is
\begin{equation}
\mathbb Y=\begin{pmatrix}
\kappa^{++}H^++\kappa^{+-}H^-\\
C_{i,z}(\kappa^{++}N^+-\kappa^{+-}N^-)\\
C_{i,x}(\kappa^{++}L^+-\kappa^{+-}L^-)\\
C_{i,y}(\kappa^{++}M^+-\kappa^{+-}M^-)\\
\kappa^{-+}H^++\kappa^{--}H^-\\
C_{i,z}(-\kappa^{-+}N^++\kappa^{--}N^-)\\
C_{i,x}(-\kappa^{-+}L^++\kappa^{--}L^-)\\
C_{i,y}(-\kappa^{-+}M^++\kappa^{--}M^-)
\end{pmatrix}.
\label{eq:app-Y}
\end{equation}
Elastic scattering conserves the total number of particles. With the phase-space convention of Eq.~\eqref{eq:app-Fmeasure}, the fixed-density condition is
\begin{align}
0&=\sum_\chi\int\frac{d^3k}{(2\pi)^3}(\mathcal D^\chi)^{-1}g^\chi_{\mathbf k}
\ \Longrightarrow\ \sum_\chi\int d\theta d\phi\,\mathcal F^\chi\Lambda^\chi=0,\\
0&=\sum_\chi\left(F^\chi d^\chi+G^\chi a^\chi+I^\chi b^\chi+J^\chi c^\chi-H^\chi\right).
\label{eq:app-conservation}
\end{align}
The common factors $eE/[(2\pi)^3\hbar]$ cancel. 
In practice, one may replace a redundant row of $\mathbb A-\mathbb I_8$ by
\begin{equation}
\mathbf c^T=(F^+,G^+,I^+,J^+,F^-,G^-,I^-,J^-),
\qquad \mathbf c^T\mathbb Z=H^++H^-,
\end{equation}
and solve the eight equations for eight unknown variables.

\section{Analytic zero-field check for a single-Weyl dispersion}
\label{app:zero_field_single_weyl}

For $N=1$, $\alpha_1=\hbar v_F$, and $B=0$, the two nodes have
the same isotropic dispersion $\epsilon_{\mathbf k}=\hbar v_F k$,
with $k_F=\epsilon_F/(\hbar v_F)$. Consequently,
$\beta^\chi=R^\chi=1$, $u_z^\chi=\cos\theta$, and the angular
weight introduced in Eq.~\eqref{eq:app-Fmeasure} reduces to
\begin{equation}
\mathcal F(\theta,\phi)=\frac{k_F^2}{v_F}\sin\theta,
\qquad
W_0=\int_0^{2\pi}d\phi\int_0^\pi d\theta\,\mathcal F
=\frac{4\pi k_F^2}{v_F}.
\end{equation}
Write $\kappa^{++}=\kappa^{--}=a_0$ and
$\kappa^{+-}=\kappa^{-+}=b_0$, with $a_0,b_0>0$.
The two valleys then have the same momentum-independent
scattering time, and the driving term is
\begin{equation}
\tau^{-1}=W_0(a_0+b_0),\qquad
h^\chi=v_F\cos\theta.
\end{equation}
The angular moments defined in Eq.~\eqref{eq:app-moments} obey
\begin{align}
F^\chi&=\tau W_0,&
O^\chi=S^\chi=V^\chi&=\frac{\tau W_0}{3},&
N^\chi&=\frac{\tau W_0v_F}{3},
\end{align}
while the first angular moments, mixed second moments, and
the remaining driving moments vanish. Rotational symmetry
and the particle-number conservation condition fix
$d^\pm=b^\pm=c^\pm=0$, and the remaining coefficients
satisfy $a^+=a^-=a$. Equation~\eqref{eq:app-coefficients}
therefore reduces to
\begin{align}
a&=t_i(a-v_F),&
t_i&=\frac{C_{i,z}(a_0-b_0)}{3(a_0+b_0)},\\
a&=-\frac{t_i v_F}{1-t_i},&
\Lambda^\chi&=-\frac{\tau v_F\cos\theta}{1-t_i}
=-\tau_{\mathrm{tr},z}^{(i)}v_F\cos\theta,
\label{eq:app-isotropic}
\end{align}
where $\tau_{\mathrm{tr},z}^{(i)}=\tau/(1-t_i)$. Thus,
\begin{align}
\left(\tau_{\mathrm{tr},z}^{(i)}\right)^{-1}
&=\frac{W_0}{3}
\left[3(a_0+b_0)-C_{i,z}(a_0-b_0)\right]\\
&=\begin{cases}
\dfrac{2W_0}{3}(a_0+2b_0),&i=0,z,\\[3pt]
\dfrac{2W_0}{3}(2a_0+b_0),&i=x,y,
\end{cases}
\label{eq:app-isotropic-tau}
\end{align}
with $C_{i,z}=+1$ for $i=0,z$ and
$C_{i,z}=-1$ for $i=x,y$.

At zero temperature, the longitudinal conductivity is
obtained from $J_z=\sigma_{zz}E$ and the linearized
distribution function as
\begin{align}
\sigma_{zz}^{(i)}(0)
&=-e^2\sum_{\chi=\pm1}\int\frac{d^3k}{(2\pi)^3}
\,v_F\cos\theta\,\Lambda^\chi\,
\delta(\epsilon_{\mathbf k}-\epsilon_F)\\
&=\frac{e^2}{(2\pi)^3\hbar}\sum_{\chi=\pm1}
\int_0^{2\pi}d\phi\int_0^\pi d\theta\,
\mathcal F\,\tau_{\mathrm{tr},z}^{(i)}
v_F^2\cos^2\theta\\
&=\frac{e^2v_F k_F^2}{3\pi^2\hbar}
\tau_{\mathrm{tr},z}^{(i)}
=\frac{e^2\epsilon_F^2}{3\pi^2\hbar^3v_F}
\tau_{\mathrm{tr},z}^{(i)},
\label{eq:app-zero-field-conductivity}
\end{align}
where both Weyl nodes have been included and
$\int_0^{2\pi}d\phi\int_0^\pi d\theta\,
\sin\theta\cos^2\theta=4\pi/3$.
Using Eq.~\eqref{eq:app-isotropic-tau} and
$W_0=4\pi k_F^2/v_F$, we finally obtain
\begin{equation}
\sigma_{zz}^{(i)}(0)=
\begin{cases}
\displaystyle\frac{e^2v_F^2}
{8\pi^3\hbar(a_0+2b_0)},&i=0,z,\\[8pt]
\displaystyle\frac{e^2v_F^2}
{8\pi^3\hbar(2a_0+b_0)},&i=x,y.
\end{cases}
\label{eq:app-zero-field-impurity}
\end{equation}
These expressions are positive for $a_0,b_0>0$ and
explicitly predict coincident zero-field conductivities
for the pairs $(\sigma_0,\sigma_z)$ and
$(\sigma_x,\sigma_y)$. For momentum-independent
point-impurity coefficients $a_0$ and $b_0$, the explicit
$\epsilon_F$ dependence cancels between the density of
states and the transport lifetime. The impurity-dependent
denominators arise directly from the coupled collision
equation; they cannot be recovered by applying impurity
signs to a scalar-channel solution after the fact.
Equation~\eqref{eq:app-zero-field-impurity} therefore
provides an analytic benchmark for the zero-field limit
of the numerical Boltzmann calculation.

\section{Velocity components for \texorpdfstring{$N=1,2,3$}{N=1,2,3}}
\label{Sec.:velocity_components}
\begin{align}
\widetilde{v}_x^{\chi}
&=
\frac{
\alpha_N^2 N\left(k_x^2+k_y^2\right)^{N-2}
}{
2\hbar
\left[
\alpha_N^2\left(k_x^2+k_y^2\right)^N
+\hbar^2k_z^2v_F^2
\right]^{5/2}
}
\Bigg[
2k_x\Bigg\{
\left(k_x^2+k_y^2\right)
\left[
\alpha_N^2\left(k_x^2+k_y^2\right)^N
+\hbar^2k_z^2v_F^2
\right]^2
\nonumber\\
&\qquad
-B\chi\eta e\,k_z N v_F\sin\gamma
\sqrt{
\alpha_N^2\left(k_x^2+k_y^2\right)^N
+\hbar^2k_z^2v_F^2
}
\left[
\alpha_N^2\left(k_x^2+k_y^2\right)^N
-\hbar^2k_z^2(N-1)v_F^2
\right]
\Bigg\}
\nonumber\\
&\qquad
-B\chi\eta e\,v_F\cos\gamma
\sqrt{
\alpha_N^2\left(k_x^2+k_y^2\right)^N
+\hbar^2k_z^2v_F^2
}
\Bigg\{
\alpha_N^2\left(k_x^2-k_y^2\right)
\left(k_x^2+k_y^2\right)^N
\nonumber\\
&\hspace{7cm}
+\hbar^2k_z^2v_F^2
\left[
k_x^2(1-2N)-k_y^2
\right]
\Bigg\}
\Bigg],
\label{eq:vx}
\\[1em]
\widetilde{v}_y^{\chi}
&=
\frac{
\alpha_N^2 k_y N\left(k_x^2+k_y^2\right)^{N-2}
}{
\hbar
\left[
\alpha_N^2\left(k_x^2+k_y^2\right)^N
+\hbar^2k_z^2v_F^2
\right]^{5/2}
}
\Bigg[
-B\chi\eta e\,k_z N v_F\sin\gamma
\sqrt{
\alpha_N^2\left(k_x^2+k_y^2\right)^N
+\hbar^2k_z^2v_F^2
}
\nonumber\\
&\qquad\times
\left[
\alpha_N^2\left(k_x^2+k_y^2\right)^N
-\hbar^2k_z^2(N-1)v_F^2
\right]
-B\chi\eta e\,k_xv_F\cos\gamma
\sqrt{
\alpha_N^2\left(k_x^2+k_y^2\right)^N
+\hbar^2k_z^2v_F^2
}
\nonumber\\
&\qquad\times
\left[
\alpha_N^2\left(k_x^2+k_y^2\right)^N
-\hbar^2k_z^2(N-1)v_F^2
\right]
+\left(k_x^2+k_y^2\right)
\left[
\alpha_N^2\left(k_x^2+k_y^2\right)^N
+\hbar^2k_z^2v_F^2
\right]^2
\Bigg],
\label{eq:vy}
\\[1em]
\widetilde{v}_z^{\chi}
&=
\frac{
v_F
}{
2\hbar\left(k_x^2+k_y^2\right)
\left[
\alpha_N^2\left(k_x^2+k_y^2\right)^N
+\hbar^2k_z^2v_F^2
\right]^2
}
\Bigg[
\alpha_N^2B\chi\eta e\,N^2\sin\gamma
\left(k_x^2+k_y^2\right)^N
\nonumber\\
&\qquad\times
\left[
\alpha_N^2\left(k_x^2+k_y^2\right)^N
-\hbar^2k_z^2v_F^2
\right]
-2\alpha_N^2B\chi\eta e\,\hbar^2k_xk_z
N v_F^2\cos\gamma
\left(k_x^2+k_y^2\right)^N
\nonumber\\
&\qquad
+2\hbar^2k_zv_F
\left(k_x^2+k_y^2\right)
\left[
\alpha_N^2\left(k_x^2+k_y^2\right)^N
+\hbar^2k_z^2v_F^2
\right]^{3/2}
\Bigg].
\label{eq:vz}
\end{align}
\end{widetext}
\bibliography{biblio.bib}
\end{document}